\documentclass[aps,preprint,preprintnumbers,nofootinbib,superscriptaddress,tightenlines,byrevtex,showpacs]{revtex4}
\usepackage{amsmath,amssymb}
\usepackage[pdftex]{graphicx}
\usepackage{comment}

\def\str{{\mathrm{str}}}

\begin{document}

\unitlength=1mm

\def\a{{\alpha}}
\def\b{{\beta}}
\def\d{{\delta}}
\def\D{{\Delta}}
\def\e{{\epsilon}}
\def\g{{\gamma}}
\def\G{{\Gamma}}
\def\k{{\kappa}}
\def\l{{\lambda}}
\def\L{{\Lambda}}
\def\m{{\mu}}
\def\n{{\nu}}
\def\w{{\omega}}
\def\O{{\Omega}}
\def\S{{\Sigma}}
\def\s{{\sigma}}
\def\t{{\tau}}
\def\th{{\theta}}
\def\x{{\xi}}

\def\ol#1{{\overline{#1}}}

\def\Dslash{D\hskip-0.65em /}
\def\dslash{{\partial\hskip-0.5em /}}
\def\vslash{{\rlap \slash v}}
\def\qbar{{\overline q}}

\def\CPT{{$\chi$PT}}
\def\QCPT{{Q$\chi$PT}}
\def\PQCPT{{PQ$\chi$PT}}
\def\tr{\text{tr}}
\def\str{\text{str}}
\def\diag{\text{diag}}
\def\order{{\mathcal O}}
\def\vit{{\it v}}
\def\vD{\vit\cdot D}
\def\am{\alpha_M}
\def\bm{\beta_M}
\def\gm{\gamma_M}
\def\smb{\sigma_M}
\def\smt{\overline{\sigma}_M}
\def\tb{{\tilde b}}
\def\tm{{\tilde m}}

\def\mc#1{{\mathcal #1}}

\def\Bbar{\overline{B}}
\def\Tbar{\overline{T}}
\def\cBbar{\overline{\cal B}}
\def\cTbar{\overline{\cal T}}
\def\pq{(PQ)}

\def\eqref#1{{(\ref{#1})}}

\newcount\hour \newcount\hourminute \newcount\minute 
\hour=\time \divide \hour by 60
\hourminute=\hour \multiply \hourminute by 60
\minute=\time \advance \minute by -\hourminute
\newcommand{\mydate}{\ \today \ - \number\hour :\number\minute}


\preprint{JLAB-THY 07-638}
\preprint{UMD 40762-385}

\title{\bf Mixed Meson Masses with Domain-Wall Valence\\
and Staggered Sea Fermions}

\author{Kostas Orginos}
\email[]{kostas@wm.edu}
\affiliation{Department of Physics, College of William and Mary, Williamsburg,
  VA 23187-8795.}
\affiliation{Jefferson Laboratory, 12000 Jefferson Avenue, 
Newport News, VA 23606.}

\author{ Andr\'e Walker-Loud} 
\email[]{walkloud@umd.edu}
\affiliation{Department of Physics, University of Maryland,
	College Park, MD 20742-4111}


\begin{abstract}
Mixed action lattice calculations allow for an additive lattice spacing dependent mass renormalization of mesons composed of one sea and one valence quark, regardless of the type of fermion discretization methods used in the valence and sea sectors.  The value of the mass renormalization depends upon the lattice actions used.  This mixed meson mass shift is an important lattice artifact to determine for mixed action calculations; because it modifies the pion mass, it plays a central role in the low energy dynamics of all hadronic correlation functions.  We determine the leading order, $\mc{O}(a^2)$, and next to leading order, $\mc{O}(a^2 m_\pi^2)$, additive mass shift of \textit{valence-sea} mesons for a mixed lattice action with domain-wall valence fermions and rooted staggered sea fermions, relevant to the majority of current large scale mixed action lattice efforts.  We find that on the asqtad improved coarse MILC lattices, this additive mass shift is well parameterized in lattice units by $\D(am)^2 = 0.034(2) -0.06(2) (a m_\pi)^2$, which in physical units, using $a=0.125$~fm, corresponds to $\D(m)^2 = (291\pm 8 \textrm{ MeV})^2 -0.06(2) m_\pi^2$.  In terms of the mixed action effective field theory parameters, the corresponding mass shift is given by $a^2 \D_\mathrm{Mix} = (316 \pm 4 \textrm{ MeV})^2$ at leading order plus next-to-leading order corrections including the necessary chiral logarithms for this mixed action calculation, determined in this work.  Within the precision of our calculation, one can not distinguish between the full next-to-leading order effective field theory analysis of this additive mixed meson mass shift and the parameterization given above.
\end{abstract}

\pacs{12.38.Gc}
\maketitle


%
%
\section{Introduction}
Recently \textit{mixed action} (or \textit{hybrid}) lattice QCD has become a commonly used method of computing hadronic observables with dynamical fermions in the chiral regime, as demonstrated by the rapid growth of mixed action calculations~\cite{Renner:2004ck,Bowler:2004hs,Bonnet:2004fr,Beane:2005rj,Edwards:2005kw,Edwards:2005ym,Beane:2006mx,Beane:2006pt,Beane:2006fk,Beane:2006kx,Alexandrou:2006mc,Beane:2006gj,Bar:2006zj,Hasenfratz:2006bq,Edwards:2006qx,Beane:2006gf}.  To compliment this progress, there have also been significant developments in our theoretical understanding of mixed action (MA) lattice QCD through the use of MA effective field theory (EFT) which describes the low energy dynamics of the theory~\cite{Bar:2002nr,Bar:2003mh,Tiburzi:2005vy,Bar:2005tu,Golterman:2005xa,Tiburzi:2005is,Chen:2005ab,Bunton:2006va,Aubin:2006hg,O'Connell:2006sh,Chen:2006wf,Walker-Loud:2006ub,Jiang:2007sn,Chen:2007ug}.  %
Mixed action QCD allows the use of a numerically cheap variant of lattice fermions in the sea sector, where a significant part of the computational time is devoted, while using valence sector fermions that respect chiral symmetry at finite lattice spacing such as domain-wall~\cite{Kaplan:1992bt,Shamir:1993zy,Furman:1994ky}  and overlap~\cite{Narayanan:1992wx,Narayanan:1994gw,Neuberger:1997fp}  fermions.  The use of these chiral fermions in the valence sector generally provides for both improved lattice spacing scaling and a simplification of the renormalization of  lattice matrix elements~\cite{Blum:1997mz,Blum:2001xb,Sasaki:2003jh,Aoki:2005ga,Orginos:2005uy}, in some cases eliminating potentially harmful power divergences.  Additionally, the EFT extrapolation formulae for observable quantities computed with MA QCD have a form very similar to those for quantities computed with chiral fermions in the valence and sea sectors, in some cases an identical form through next-to-leading order (NLO)~\cite{Chen:2006wf,Chen:2007ug}, which have only small deviations from the continuum chiral perturbation theory formulae~\cite{Weinberg:1966kf,Gasser:1983yg,Gasser:1984gg}. 
In other words mixed action lattice QCD provides most of the benefits of lattice fermions with chiral symmetry, while avoiding the large cost of using these fermions in the sea sector.

The most popular mixed action scheme was developed by LHPC~\cite{Renner:2004ck,Edwards:2005kw} and has been used in several recent calculations~\cite{Renner:2004ck,Bonnet:2004fr,Beane:2005rj,Edwards:2005kw,Edwards:2005ym,Edwards:2006qx,Beane:2006mx,Beane:2006pt,Beane:2006fk,Beane:2006kx,Alexandrou:2006mc,Beane:2006gj,Beane:2006gf}. In this approach domain-wall fermions are used in the valence sector while the fermionic determinant is represented by  asqtad-improved~\cite{Orginos:1998ue,Orginos:1999cr} fermions using the  publicly available MILC configurations~\cite{Bernard:2001av}.  
The MILC configurations are generated with rooted staggered sea fermions to recover the correct number of degrees of freedom in the continuum limit.  There has recently been much discussion regarding the validity of the rooting procedure~\cite{Durr:2004ta,Bernard:2006zw,Creutz:2006ys,Bernard:2006vv,Durr:2006ze,Hasenfratz:2006nw,Bernard:2006ee,Shamir:2006nj,Bernard:2006qt,Creutz:2007yg,Creutz:2007pr} and a comprehensive review of the issues can be found in Ref.~\cite{Sharpe:2006re}. 

Independent of these issues related to rooted staggered fermions, mixed action calculations come with their own set of unique challenges.  At finite lattice spacing, MA theories violate unitarity.  Also, hadrons composed of all valence quarks, all sea quarks, or a mix of valence and sea quarks each receive different lattice spacing dependent mass renormalizations.  However, these issues are addressed in a MAEFT framework~\cite{Bar:2002nr} which is a natural extension of partially quenched chiral perturbation theory (PQ$\chi$PT)~\cite{Bernard:1993sv,Sharpe:1997by}.  The unitarity violations in a mixed action effective theory are identical in form to those in partially quenched effective theories which is transparent with the introduction of \textit{partial quenching parameters}~\cite{Chen:2005ab,Chen:2006wf}: these provide the means to quantify the unitarity violations at the EFT level.  

One particularly nice feature of MA theories is that the form of the leading order (LO) lattice spacing dependent additive mass renormalization for a mixed meson composed of one valence and one sea quark is universal, \textit{i.e.} independent of the particular lattice actions used for the valence and sea sectors, and is described by one parameter, known in the literature as $C_\mathrm{Mix}$~\cite{Bar:2005tu}.  This parameter, although not physical, is an important lattice artifact to be determined for MA calculations; the long-range correlations of hadronic observables are modified by this parameter as the infrared behavior is dominated by the lightest particles in the system, which include the mixed valence-sea mesons.  There are some observables in which it has been shown this parameter does not appear even through NLO in the MA expansion~\cite{Chen:2005ab,Aubin:2006hg,Chen:2006wf}, although this is generally not the case~\cite{Chen:2007ug}.

\bigskip
In this article we present our computation of the lattice spacing shift to the mixed pion mass, specific to domain-wall valence fermions and the rooted staggered coarse MILC fermions, relevant for the most commonly performed mixed action calculations at present.  We perform this calculation by measuring the long-Euclidean time behavior of two-point correlation functions constructed with one domain-wall quark and one staggered quark, tied together to form mixed pions.  As with all calculations using rooted staggered sea fermions, the validity, and perhaps usefulness of this work depends upon the the rooting procedure common to staggered lattice actions.  By calculating the mixed pion mass, we are able to determine a numerical value of the MAEFT low energy constant, $C_\mathrm{Mix}$.%
\footnote{We can unambiguously determine the LO mixed pion mass shift, $a^2 \D_\mathrm{Mix} = a^2 16 C_\mathrm{Mix} / f^2$, however the extracted value of $C_\mathrm{Mix}$ will obviously depend on the choice of $f$, whether it is chosen at a given pion mass, or in the chiral but not continuum limit, or in both limits, etc.} 
We find that the LO description of the mixed meson mass shift in the MAEFT is insufficient to describe our computational results.  Therefore, in Sec.~\ref{sec:MALQCD}, we determine NLO extrapolation formula for a mixed meson with chiral valence fermions and rooted staggered sea fermions.  In Sec.~\ref{sec:CalcDetails} we present the details of our lattice calculation, highlighting some peculiarities which arise in the calculation of our mixed domain-wall--rooted-staggered meson correlation functions (the details are collected in the appendix).  In Sec.~\ref{sec:results}, we analyze our results with the NLO extrapolation formula, finding good agreement.  Additionally, we find a simple quadratic in $m_\pi^2$ parameterization of the mixed pion mass splitting provides an equally good description of this quantity, allowing for a user-friendly consumption of our results.  We then conclude in Sec.~\ref{sec:concl}.

%
%
\section{Mixed Meson masses to NLO in MAEFT \label{sec:MALQCD}}

We begin with a brief introduction of mixed action effective field theory to set our conventions and motivate our study of the additive mixed meson mass renormalization.  The EFTs appropriate to describe the low-energy dynamics of mesons were first developed for Ginsparg-Wilson valence fermions and Wilson sea fermions in Ref.~\cite{Bar:2002nr,Bar:2003mh} and for staggered sea fermions in Ref.~\cite{Bar:2005tu}.  As mentioned above, these theories are a natural extension of partially quenched chiral perturbation theory (PQ$\chi$PT)~\cite{Bernard:1993sv,Sharpe:1997by} (with PQ$\chi$PT recovered in the continuum limit, $a \rightarrow 0$, of MAEFT).  The MAEFT is a dual expansion in the quark mass and the lattice spacing, as are all lattice EFTs, and is constructed in a two step process developed by Sharpe and Singleton~\cite{Sharpe:1998xm}: one first constructs the Symanzik effective continuum theory for the corresponding lattice action~\cite{Symanzik:1983gh,Symanzik:1983dc} and then one constructs the low energy effective theory corresponding to this Symanzik action.

For the coarse MILC lattices, the LO lattice spacing effects, which begin at $\mc{O}(g^2a^2)$, are comparable to the LO quark mass effects, as seen by the splitting amongst the various staggered taste meson masses~\cite{Aubin:2004fs}, which suggests a useful power counting is $\varepsilon_{a}^2 \sim \varepsilon_{m}^2$ where
\begin{equation}
	\varepsilon_a^2 \sim a^2 \Lambda_{QCD}^2\quad, \quad 
	\varepsilon_m^2 \sim \frac{m_q}{\Lambda_{QCD}}\, .
\end{equation}

At leading order in this power counting, the MA Lagrangian with GW valence quarks and any type of sea quarks is given by~\cite{Bar:2003mh,Bar:2005tu}
\begin{equation}\label{eq:PQChPT}
	\mc{L} = \frac{f^2}{8} \str \left( \partial_\mu \Sigma \partial^\mu \Sigma^\dagger \right)
		- \frac{f^2 B_0}{4} \str \left( m_q \Sigma^\dagger + \Sigma m_q^\dagger \right)
	+a^2 \Big( \mc{U}_{sea} - \mc{U}_{vs} \Big)\, ,
\end{equation}
where we use the normalization $f \simeq 132$~MeV, and the sigma field contains the mesons, 
\begin{equation}
	\S = \textrm{exp} \left( \frac{2 i \Phi}{f} \right) ,  \;\;\; \\
	\Phi = \left( \begin{array}{cc}
		M & \chi^\dagger\\
		\chi & \tilde{M}
		\end{array} \right) \, .
\label{eq:sigma}
\end{equation}
The matrices $M$ and $\tilde{M}$ contain bosonic mesons while $\chi$ and $\chi^\dagger$ contain fermionic mesons with one ghost quark or antiquark.  For an explicit representation see for example Ref.~\cite{Chen:2006wf}.  

In this article, we are interested in staggered sea fermions, for which the LO potential can be found in Refs.~\cite{Lee:1999zx,Aubin:2003mg}.  The mixed meson potential has a universal form, regardless of the type of sea and valence fermions, and is given by 
\begin{equation}
	\mc{U}_{vs} = C_\mathrm{Mix}\, \str \left( T_3 \S T_3 \S^\dagger \right)\, ,
\end{equation}
where the flavor matrix $T_3$ is a difference in projectors onto the valence and sea sectors of the theory,
\begin{equation}
	T_3 = \mc{P}_S -\mc{P}_V = \textrm{diag} (-\mathbb{I}_V, \mathbb{I}_S, -\mathbb{I}_V)\, ,
\end{equation}
where $\mathbb{I}_V$ and $\mathbb{I}_S$ are identity matrices which span the space of valence and sea fermions respectively. %
These mixed and sea potentials, $\mc{U}_{vs}$ and $\mc{U}_{sea}$ give rise to the LO additive lattice spacing dependent renormalization of the meson masses.  The sea-potential splits the staggered meson spectrum into five different tastes at this order while the mixed-potential provides a flavor and taste independent mass shift to all \textit{valence-sea} mesons composed of one valence quark, $v$, and one sea quark, $s$, giving rise to the LO parameterization of the spectrum
\begin{align}\label{eq:LOMasses}
	m_{v_1 v_2}^2 &= B_0 (m_{v_1} +m_{v_2} )\, ,\nonumber\\
	\tilde{m}_{vs}^2 &= B_0 (m_v +m_s ) +a^2 \Delta_\mathrm{Mix}\, ,\nonumber\\
	\tilde{m}_{s_1 s_2, B}^2 &= B_0 (m_{s_1} +m_{s_2} ) +a^2 \Delta_B\, .
\end{align}
We are adopting the convention of using \textit{tildes} to denote masses which receive additive lattice spacing dependent corrections~\cite{Chen:2005ab}.  The various taste splitting ($a^2 \Delta_B$) have been computed on the coarse (and fine) MILC lattices and are found to be~\cite{Aubin:2004fs},
\begin{align}\label{eq:tasteSplit}
	a^2 \D_5 &= 0 \, ,\nonumber\\
	a^2 \D_{A} &\simeq (276 \pm 1 \textrm{ MeV})^2 \, ,\nonumber\\
	a^2 \D_{T} &\simeq (348 \pm 2 \textrm{ MeV})^2 \, ,\nonumber\\
	a^2 \D_{V} &\simeq (404 \pm 2 \textrm{ MeV})^2 \, ,\nonumber\\
	a^2 \D_I &\simeq (446 \pm 6 \textrm{ MeV})^2 \, .
\end{align}
The mixed meson mass correction,
\begin{equation}\label{eq:CMix}
	a^2 \Delta_\mathrm{Mix} = a^2 \frac{16\, C_\mathrm{Mix}}{f^2}\, ,
\end{equation}
has not been previously determined, and it is this quantity in particular that we have computed.%
\footnote{In Ref.~\cite{Bunton:2006va}, this mass shift was estimated from a MA lattice calculation of the pion form factor using domain-wall valence and the coarse MILC sea quarks~\cite{Bonnet:2004fr} by using the appropriate MA extrapolation formula.  Unfortunately, only one lattice numerical result was in the chiral regime, and so only a crude estimate was made, $a^2 \D_\mathrm{Mix} \sim (380 \textrm{ MeV})^2$.} 
This is one of the more important unphysical parameters (lattice artifacts) in a given MA lattice simulation, because it modifies the mass of the mixed mesons, and therefore modifies the chiral behavior of all hadronic correlation functions, generally at NLO in the MAEFT extrapolation formulae~\cite{Chen:2007ug} (for some observables this dependence does not appear until at least NNLO~\cite{Chen:2005ab,Aubin:2006hg,Chen:2006wf}).

To isolate this mixed meson mass renormalization we calculated the quantity
\begin{equation}\label{eq:Dmsqr}
\D m^2 = \tilde{m}_{vs}^2 -\frac{1}{2} (m_{vv}^2 + m_{ss,5}^2)\, .
\end{equation}
The method to calculate the mixed meson mass renormalization is not unique, however this particular choice has a few distinct advantages.  First, as is obvious from Eq.~\eqref{eq:LOMasses}, the LO quark mass dependence exactly cancels in this mass difference and thus the dominant contribution to this mass difference will be precisely the quantity of interest, $a^2 \D_\mathrm{Mix}$.  Second, this choice is entirely unphysical not just at LO but to all orders in MAEFT: it must vanish in the continuum QCD limit.  In particular, for degenerate valence and sea quark masses, this choice must be proportional to the square of the lattice spacing, and therefore is a clean and good way to directly determine the size of the MA lattice spacing artifacts.

As we will show in Section~\ref{sec:results}, we find that there is non-trivial quark mass dependence in this mass splitting signaling a sensitivity to NLO effects.  Therefore we not only determine the LO mass renormalization but also the NLO mass renormalization of the mixed valence sea mesons.  This requires the NLO extrapolation formulae for all three meson masses, the valence-valence~\cite{Bar:2005tu}, the sea-sea~\cite{Aubin:2003mg} and the mixed meson.  The NLO MA formula for the mixed valence-sea fermions has not been worked out before.  It is a straightforward computation, similar to computing loop graphs in MAEFT as well as a staggered $\chi$PT~\cite{Lee:1999zx,Aubin:2003mg,Aubin:2003uc,VandeWater:2005uq,Bailey:2007iq}.  However, as this is not the focus of this work, we give here just one intermediate step and the answer.  To be precise, we shall give the answer for a mixed meson mass composed of a domain-wall valence \textit{up}--quark $u$, and a staggered sea \textit{down}--quark $l$ of arbitrary quark-taste $t$ (not to be confused with meson-taste $B$).  However, it will be clear how to generalize our expression for an arbitrary flavor non-singlet composition.  First we provide an integral expression for the NLO loop correction to the mixed meson mass (as a Minkowski space momentum integral with regularization scheme $R$),
\begin{align}\label{eq:mvsNLOIntegral}
\d\tilde{m}_{ul}^2 =&\ \frac{1}{3f^2} \int_R \frac{d^4 k}{(2\pi)^4} \bigg\{
		(k^2 - m_\pi^2) \mc{G}_{\eta_u\eta_u}^{C+D}(k^2)
		-2(k^2 + 2 \tm_{ul}^2) \mc{G}_{\eta_l \eta_u}^D(k^2)
		+(k^2 - \tilde{m}_{ll,I}^2) \mc{G}_{\eta_l\eta_l,I}^D(k^2)
	\nonumber\\&\ 
		+a^2 \D_I\, \mc{G}_{\eta_l\eta_l,I}^D(k^2)
		-6 a^2\D_\mathrm{Mix} \frac{1}{4} \sum_{t=1}^4  \sum_{F_{s}=j,l,r} \mc{G}_{uF_s,t}(k^2)
		+\frac{1}{4}\sum_B \sum_{F_{s}=j,l,r}a^2 \D_B\, \mc{G}_{lF_s,B}^C(k^2)
	\nonumber\\&\ 
		+4(k^2 -m_{ll,5}^2)\mc{G}_{\eta_l\eta_l,V}^D(k^2)
		+4(k^2 -m_{ll,5}^2)\mc{G}_{\eta_l\eta_l,A}^D(k^2)
	\bigg\}\, .
\end{align}
The first line of this expression contains all the physical contributions which survive the continuum PQ and QCD limits, while the other two lines contain only \textit{unphysical} MA and staggering lattice artifacts.  The flavor--diagonal momentum space propagators contain both ``connected" (C) as well as ``disconnected" (D) or haripin pieces, and standard expressions for these propagators can be found for example in Ref.~\cite{Sharpe:2000bc}.  The meson-taste labels on the propagators, $B = \{5,A,T,V,I\}$ denote that the sea-sea meson masses in these expressions are of the taste $B$, with the mass renormalizations given in Eq.~\eqref{eq:tasteSplit}.  The expressions for the disconnected taste--$V$ and taste--$A$ propagators are slightly different from the PQ form and can be found in Ref.~\cite{Aubin:2003mg}.  The explicit factors of $\frac{1}{4}$ in front of the taste sums are necessary to account for the ``rooting" of the staggered fermion determinant and can be understood either with the quark-flow picture~\cite{Sharpe:1992ft} or with the replica method~\cite{Damgaard:2000gh,Bernard:2006zw}.  The summation over the four quark-tastes is denoted by a sum over $t$ while the summation over the 16 meson-tastes is denoted as a sum over $B$, and the explicit flavor sums run over the sea-quark flavors, denoted $j,l,r$ following the convention set in Ref.~\cite{Chen:2001yi}.  Lastly, to determine the expression for a mixed meson of valence flavor $V$ and sea flavor $S$, one simply makes the replacements in Eq.~\eqref{eq:mvsNLOIntegral} $u \rightarrow V$ and $l \rightarrow S$, and then computes the given integrals in Eq.~\eqref{eq:mvsNLOIntegral}.

We now provide the expression for the mixed pion mass to NLO in MAEFT of flavor $ul$.  The NLO MA formula for the ``charged" mixed pion mass in the degenerate light quark limit is given by (using dimensional regularization and the modified minimal subtraction scheme conventional to $\chi$PT~\cite{Gasser:1983yg,Gasser:1984gg}.)
\begingroup
\small
\begin{align}\label{eq:mvsNLO}
\tm_{ul}^2 =&\ B_0 (m_u +m_l) +a^2 \Delta_\mathrm{Mix} 
	+\frac{16(B_0(m_u+m_l))^2}{f_\pi^2} \Big(2L_8(\mu) -L_5(\mu) \Big)
	\nonumber\\&\ 
	+B_0(m_u+m_l) \left[ \frac{32B_0 (2m_l +m_r)}{f_\pi^2} \Big(2L_6(\mu) - L_4(\mu) \Big)
		+a^2 L_{a^2m}(\mu)
	\right]
	+a^4 L_{a^4}(\mu)
	\nonumber\\&\ 
	+\frac{m_{ul}^2}{(4\pi f_\pi)^2} \left[ 
		\frac{1+\frac{2}{3}(\tilde{\D}_{lu}^2 / \D_{Xl}^2)}{1+(\tilde{\D}_{lu}^2 / \D_{Xl}^2)}\, 
			m_{\pi}^2 \ln \left( \frac{m_\pi^2}{\mu^2} \right) 
	-\frac{\tm_{X_I}^2}{3(1+(\tilde{\D}_{lu}^2 / \D_{Xl}^2))} 
		\ln \left( \frac{\tm_{X_I}^2}{\mu^2} \right) 
	\right]
	\nonumber\\&\ 
	-\frac{2a^2 \D_\mathrm{Mix}}{(4\pi f_\pi)^2} \left[
		2\tm_{ul}^2 \ln \left( \frac{\tm_{ul}^2}{\mu^2} \right)
		+\tm_{ur}^2 \ln \left( \frac{\tm_{ur}^2}{\mu^2} \right)
	\right]
	\nonumber\\&\ 
	-\frac{a^2\D_I}{6(4\pi f_\pi)^2} \left[ \tm_{ll,I}^2 \ln \left( \frac{\tm_{ll,I}^2}{\mu^2} \right)
		-\frac{\tm_{X,I}^2}{3}  \ln \left( \frac{\tm_{X,I}^2}{\mu^2} \right)
	\right]
	\nonumber\\&\ 
	+\frac{4 a^2 \d_V^\prime}{3(4\pi f_\pi)^2} \bigg[
		\frac{(\tm_{rr_V}^2 -\tm_{\eta_V^\prime}^2)(\tm_{\eta_V^\prime}^2 -m_{\pi_5}^2) 
			\tm_{\eta_V^\prime}^2}
			{(\tm_{\eta_V^\prime}^2 -\tm_{\eta_V}^2)(\tm_{\eta_V^\prime}^2 -\tm_{\pi_V}^2)}
			\ln \left( \frac{\tm_{\eta_V^\prime}^2}{\mu^2} \right)
	\nonumber\\&\ 
		+\frac{\D_{rj}^2\, a^2\D_V\, \tm_{\pi_V}^2}
			{(\tm_{\eta_V^\prime}^2 -\tm_{\pi_V}^2)(\tm_{\eta_V}^2 -\tm_{\pi_V}^2)}
			\ln \left( \frac{\tm_{\pi_V}^2}{\mu^2} \right)
		-\frac{(\tm_{rr_V}^2 -\tm_{\eta_V}^2)(\tm_{\eta_V}^2 -m_{\pi_5}^2) \tm_{\eta_V}^2}
			{(\tm_{\eta_V^\prime}^2 -\tm_{\eta_V}^2)(\tm_{\eta_V}^2 -\tm_{\pi_V}^2)}
			\ln \left( \frac{\tm_{\eta_V}^2}{\mu^2} \right)
	\bigg]
	\nonumber\\&\ 
	+\frac{4 a^2 \d_A^\prime}{3(4\pi f_\pi)^2} \bigg[
		\frac{(\tm_{rr_A}^2 -\tm_{\eta_A^\prime}^2)(\tm_{\eta_A^\prime}^2 -m_{\pi_5}^2) 
			\tm_{\eta_A^\prime}^2}
			{(\tm_{\eta_A^\prime}^2 -\tm_{\eta_A}^2)(\tm_{\eta_A^\prime}^2 -\tm_{\pi_A}^2)}
			\ln \left( \frac{\tm_{\eta_A^\prime}^2}{\mu^2} \right)
	\nonumber\\&\ 
		+\frac{\D_{rj}^2\, a^2\D_A\, \tm_{\pi_A}^2}
			{(\tm_{\eta_A^\prime}^2 -\tm_{\pi_A}^2)(\tm_{\eta_A}^2 -\tm_{\pi_A}^2)}
			\ln \left( \frac{\tm_{\pi_A}^2}{\mu^2} \right)
		-\frac{(\tm_{rr_A}^2 -\tm_{\eta_A}^2)(\tm_{\eta_A}^2 -m_{\pi_5}^2) \tm_{\eta_A}^2}
			{(\tm_{\eta_A^\prime}^2 -\tm_{\eta_A}^2)(\tm_{\eta_A}^2 -\tm_{\pi_A}^2)}
			\ln \left( \frac{\tm_{\eta_A}^2}{\mu^2} \right)
	\bigg]
	\nonumber\\&\ 
	+\frac{1}{12(4\pi f_\pi)^2} \sum_{F_s=j,l,r} \bigg[
		a^2 \D_I\, \tm_{lF_s,I}^2 \ln \left( \frac{\tm_{lF_s,I}^2}{\mu^2} \right)
		+4a^2 \D_A\, \tm_{lF_s,A}^2 \ln \left( \frac{\tm_{lF_s,A}^2}{\mu^2} \right)
	\nonumber\\&\ 
		+4a^2 \D_V\, \tm_{lF_s,V}^2 \ln \left( \frac{\tm_{lF_s,V}^2}{\mu^2} \right)
		+6a^2 \D_T\, \tm_{lF_s,T}^2 \ln \left( \frac{\tm_{lF_s,T}^2}{\mu^2} \right)
	\bigg]\, .
\end{align}
\endgroup
In this expression, the mass of the sea-sea $\eta$-octet field, which is of the taste-$I$ variety, is denoted $\tilde{m}_{X_I}^2$.  The light quark partial quenching parameter is $\tilde{\D}_{lu}^2 = \tilde{m}_{ll,I}^2 - m_\pi^2$.  It turns out to be convenient to express the $SU(3)$ breaking in the sea-quark masses in two ways,
\begin{equation}
	\D_{Xl}^2 = \frac{2}{3}\D_{rj}^2 = \frac{2}{3} \Big( \tm_{rr,B}^2 - \tm_{jj,B}^2 \Big)\, .
\end{equation}
The new vector masses, $\tm_{\eta_V}$ and $\tm_{\eta_V^\prime}$ are the eigenvalues of a mass matrix composed of the quark mass contributions as well as the vector hairpin coupling $a^2 \d_V^\prime$ contributions to these masses, for which the explicit form can be found in Eqs. (58) and (59) of Ref.~\cite{Aubin:2003mg}, and similarly for the new axial meson masses and hairpin coupling as well.

We now provide the NLO extrapolation formulae needed for our work with the mass tunings employed in most MA calculations to date~\cite{Renner:2004ck,Bonnet:2004fr,Beane:2005rj,Edwards:2005kw,Edwards:2005ym,Beane:2006mx,Beane:2006pt,Beane:2006fk,Beane:2006kx,Alexandrou:2006mc,Beane:2006gj,Edwards:2006qx,Beane:2006gf}, in which the domain-wall pion mass is tuned to the Goldstone staggered pion mass to within a few percent accuracy.  This means we can treat all the quark masses and partial quenching parameters as equal,
\begin{align}
	m_u = m_d &= m_j = m_l\, ,
	\nonumber\\
	m_r &= m_s\, ,
	\nonumber\\
	\tilde{\D}_{ju}^2 = \tilde{\D}_{lu}^2 &= \tilde{\D}_{rs}^2 = a^2 \D_I\, ,
\end{align}
The mixed meson mass renormalization defined in Eq.~\eqref{eq:Dmsqr} is given through NLO by
\begin{align}\label{eq:DmvsNLO}
\D m^2 =&\ \Big( a^2 \Delta_\mathrm{Mix} 
	+a^4 L_{a^4}(\mu) \Big)
	+m_{\pi}^2 a^2 L_{a^2m}(\mu)
	\nonumber\\&\ 
	-\frac{4a^2 \D_\mathrm{Mix}}{(4\pi f_\pi)^2} 
		(m_{\pi}^2 +a^2\D_\mathrm{Mix}) \ln \left( \frac{m_{\pi}^2 +a^2\D_\mathrm{Mix}}{\mu^2} \right)
	\nonumber\\&\ 
	-\frac{2a^2 \D_\mathrm{Mix}}{(4\pi f_\pi)^2}
		(m_{K}^2 +a^2\D_\mathrm{Mix}) \ln \left( \frac{m_{K}^2 +a^2\D_\mathrm{Mix}}{\mu^2} \right)
	\nonumber\\&\ 
	+\chi^{NLO}\, ,
\end{align}
where $\chi^{NLO}$ is a lengthy expression containing most of the chiral logarithms determined by Eqs.~\eqref{eq:mvsNLO}, \eqref{eq:DmvsNLO}, Eq.~(75) of Ref.~\cite{Aubin:2003mg} and for example Eq.~(B1) of Ref.~\cite{Chen:2006wf}.  We see from Eq.~\eqref{eq:DmvsNLO} that the mixed meson mass renormalization depends upon the LO mass renormalization $a^2\D_\mathrm{Mix}$ linearly, quadratically and also logarithmically.  The other important thing to note is that $\chi^{NLO}$ is a known function given the inputs of staggered lattice artifacts determined by MILC~\cite{Aubin:2004fs} and that all the dependence upon the physical counterterms has exactly cancelled.%
\footnote{This is only true for degenerate sea and valence quark masses, otherwise this mixed meson mass renormalization depends upon one linear combination of Gasser-Leutwyler constants, the counterterms of $\chi$PT~\cite{Weinberg:1966kf,Gasser:1983yg,Gasser:1984gg}, 
$\D m^2 \propto \frac{16(B_0(m_l-m_u))^2}{f^2} \left(L_5 -2L_8 \right)$.} 
Therefore, this function only depends upon three unknown terms, $a^2\D_\mathrm{Mix}$, $a^4 L_{a^4}(\mu)$ and $a^2 L_{a^2m}(\mu)$.  With only one lattice spacing we can not independently determine $a^2\D_\mathrm{Mix}$ and $a^4 L_{a^4}(\mu)$ and so we treat these as one parameter in our fits which we report in Section~\ref{sec:results}.  We now turn to the lattice calculations performed for this work before presenting the results of our analysis.

%
%
\section{Lattice Calculation\label{sec:CalcDetails}}
%
%
\subsection{Lattice actions and parameters\label{sec:LattParams}}

Here we are computing the additive mixed meson mass renormalization relevant to  the lattice actions used in a number of recent mixed action lattice computations~\cite{Bonnet:2004fr,Beane:2005rj,Edwards:2005kw,Edwards:2005ym,Beane:2006mx,Beane:2006pt,Beane:2006fk,Beane:2006kx,Alexandrou:2006mc,Beane:2006gj,Beane:2006gf}. These calculations use  domain-wall valence quarks with $N_f = 2+1$ asqtad-improved~\cite{Orginos:1998ue,Orginos:1999cr} quarks included in the publicly available MILC configurations~\cite{Bernard:2001av}. In the valence sector, HYP-smearing~\cite{Hasenfratz:2001hp,DeGrand:2002vu,DeGrand:2003in,Durr:2004as} of the gauge field entering the domain-wall fermion Dirac operator has been used  in order to reduce the explicit chiral symmetry breaking present at finite 5th dimensional extent.  We worked at a lattice spacing  $a\sim 0.125$ fm keeping the strange quark mass fixed near its physical value, $am_s = 0.05$ and used a range of degenerate light quark masses. See Table~\ref{tab:LattParameters} for details.  The domain-wall propagators we used were generated  by the LHP  and NPLQCD Collaborations using Dirichlet boundary conditions (DBC) and reducing the time extent of the original MILC lattices from 64 to 32.  To study the systematics introduced by Dirichlet boundary conditions we additionally used a set of full volume anti-periodic domain-wall propagators generated on 171 of the m010 lattices.   In all cases  the domain-wall fermion action had an extra dimension of $L_s = 16$ and a domain-wall height of $M_s = 1.7$.  The domain-wall fermion quark mass  was tuned such that the domain-wall pion mass is equal to the staggered Goldstone pion mass  within a few percent~\cite{Renner:2004ck,Edwards:2005kw}.  To construct the mixed mesons, we computed staggered propagators on these lattices with masses set equal to the corresponding sea quark masses.  For computational simplicity, we used the same boundary conditions for the staggered fermions as was used on the domain-wall and fermions.  All propagator calculations and meson interpolating operators were created with the Chroma software suite~\cite{Edwards:2004sx}.

%
%
\begin{table}[t]
\caption{\label{tab:LattParameters} We lists the parameters of the MILC gauge configurations and domain-wall propagators we used in this work.  With the mass m010 lattices, we used propagators with both anti-periodic boundary conditions as well as Dirichlet boundary conditions.  For this work, we generated an equal number of staggered propagators with masses equal to those in the configurations.}
\begin{ruledtabular}
\begin{tabular}{cccccc}
Ensemble & $am_l$ & $am_s$ & $am_l^{dwf}$ & $10^3 \times am_{res}\footnote{Determined by the LHP Collaboration.}$ & \# propagators \\
\hline
2064f21b676m010m050 & 0.010 & 0.050 & 0.0138 & $1.552 \pm 0.027$ & $171$ Full--V \\
2064f21b676m010m050 & 0.010 & 0.050 & 0.0138 & $1.552 \pm 0.027$ & 447 DBC \\
2064f21b679m020m050 & 0.020 & 0.050 & 0.0313 & $1.239 \pm 0.028$ & 222 DBC \\
2064f21b681m030m050 & 0.030 & 0.050 & 0.0478 & $0.982 \pm 0.030$ & 564 DBC \\
2064f21b683m040m050 & 0.040 & 0.050 & 0.0644 & $0.834 \pm 0.024$ & 349  DBC \\
2064f3b685m050 & 0.050 & 0.050 & 0.0810 & $0.726 \pm 0.030$ & 416 DBC
\end{tabular}
\end{ruledtabular}
\end{table}

%
%
\subsection{Meson interpolating operators and two-point correlation functions\label{sec:LattCorrelators}}

In order to compute the mixed meson mass renormalization defined in Eq.~\eqref{eq:Dmsqr}, we use a source for each of the domain-wall pion, the staggered Goldstone pion and the mixed domain-wall staggered pion.  For the domain-wall pion we use the standard pion interpolating operator
\begin{equation}
	\pi(x) = \bar{\psi}_{dw}(x)\, \g_5\, \psi_{dw}(x)\, .
\end{equation}
To create the staggered pion and the mixed mesons, we use a ``Wilsonized" staggered quark,
\begin{equation}
	\psi_{s}^\a(x) = \left( \Omega(x)\, C \right)^\a\, \chi(x)\, ,
\end{equation}
where $\a$ is a Dirac-spinor index, $C$ can be chosen as a constant Dirac-spinor, $\chi(x)$ is the single component staggered fermion field and
\begin{equation}
	\Omega(x) = \prod_{\mu=1}^{4} (\g_\mu)^{x_\mu / a}\, ,
\end{equation}
is the Kawamoto-Smit multiplicative phase factor~\cite{Kawamoto:1981hw} which accomplishes the Susskind spin-diagonalization~\cite{Kogut:1974ag,Banks:1975gq,Susskind:1976jm} relating staggered fermions to naive fermions, and we are using Euclidean Hermitian gamma matrices,
\begin{equation}\label{eq:EuclidCommute}
	\{ \g_\mu, \g_\nu \} = 2 \d_{\mu\nu}\, .
\end{equation}
This allowed us to use the existing Chroma software supplemented by a minimal amount of new software to create the staggered propagators as well as the staggered meson and mixed meson source and sink operators.  
For example, the staggered Goldstone pion can be created with the interpolating operator
\begin{align}
	\pi_5(x) &= \bar{\psi}_s(x)\, \g_5\, \psi_s(x) 
	\nonumber\\&
	\propto \e(x)\, \bar{\chi}(x)\, \chi(x)\, ,
\end{align}
which is known to have the quantum numbers of a spin zero pseudo-scalar meson~\cite{Golterman:1985dz} and the phase is $\e(x) = (-1)^{\sum_\mu (x_\mu / a)}$.
In this work, we not only look at the mixed pseudo-scalar correlator created with the interpolating operator,
\begin{equation}\label{eq:pivs}
	\pi_{vs}(x) = \bar{\psi}_{dw}(x)\, \g_5\, \psi_s(x)\, ,
\end{equation}
and its Hermitian conjugate, but we also study the two-point correlation functions of all the mixed meson sources created with the interpolating operators
\begin{align}
	\pi_{vs}^\Gamma(x) &= \bar{\psi}_{dw}(x)\, \Gamma\, \psi_s(x)\, , \label{eq:pivsArbitrary}
	\\& \textrm{with}
	\nonumber\\
	\Gamma =&\ \{ \mathbb{I}, \g_5, \g_\mu, \g_5 \g_\mu, \g_\mu \g_\nu \}\, , \label{eq:GammaMvs}
\end{align}
and $\mu \neq \nu$.  The reasons for this will become clear below.  This construction is very similar to that used for heavy mesons with staggered light quarks~\cite{Wingate:2002fh}.  There is a point we should note here.  The virtual mixed mesons which propagate in internal loops of of mixed action hadronic correlation functions are composed of one domain-wall or overlap valence fermion, and one rooted staggered sea fermion.  To determine the mixed meson mass, we construct a correlation function with a domain-wall fermion and a non-rooted staggered fermion.  One might be concerned that the non-locality of the rooted staggered action~\cite{Bernard:2006ee} invalidates this method of determining the mixed meson mass.  However, we need make no more assumptions about the vanishing non-localities in the continuum limit than are necessary for the construction and analysis of any correlation function with staggered valence fermions on staggered sea fermions.  For example, this is how the MILC Collaboration determines the masses of the various taste mesons which propagate in virtual loops of correlation functions constructed with only the taste-5 mesons~\cite{Aubin:2004fs}.  Furthermore, it is in this sense that a staggered lattice action in which the valence and sea quark masses are degenerate is still a partially quenched theory~\cite{Bernard:2006zw}.

The zero momentum, two point correlation function created by the domain-wall pion source is given by%
\footnote{Higher energy states in the sum will not necessarily correspond to single particle mass eigenstates but as we are only interested in the ground state we use $m_n$ instead of $E_n$ to represent the energy of a given mode in the sum.} 
\begin{align}
C_{\pi}(t) &= \sum_{\mathbf{x}} \langle\, \pi(\mathbf{x},t)\, \pi^\dagger(\mathbf{0},0)\, \rangle
	\nonumber\\&
	= \sum_n A_n \textrm{cosh}\Big( m_n(t-N_T/2) \Big)\, ,
\end{align}
on the full periodic volume%
\footnote{Anti-periodic boundary conditions for the quarks gives rise to periodic  boundary conditions (PBC) for pions.} 
and on the lattices with Dirichlet boundary conditions with the replacement $\cosh \rightarrow \exp$.
The staggered meson interpolating operator creates both states which are ``straight" and those which oscillate in time, given by
\begin{align}
C_{\pi_5}(t) &= \sum_{\mathbf{x}} \langle\, \pi_5(\mathbf{x},t)\, \pi_5^\dagger(\mathbf{0},0)\, \rangle
	\nonumber\\&
	= \sum_n A_n \textrm{cosh}\Big( m_n(t-N_T/2) \Big)
		+\sum_{n^\prime} B_{n^\prime} (-1)^{n_t} \textrm{cosh}\Big( m_{n^\prime}(t-N_T/2) \Big)\, ,
\end{align}
with $n_t = t/a$ (and $\cosh \rightarrow \exp$ for DBC).
For the staggered states created with the staggered meson interpolating field, the states which oscillate in time also have opposite intrinsic parity as compared to the straight states~\cite{Golterman:1985dz}, and are therefore significantly heavier than the lightest Goldstone pion.

The mixed meson correlation functions also contain states which have a time-oscillatory behavior.  The mixed meson correlation functions created with the arbitrary mixed meson interpolating operators, Eq.~\eqref{eq:pivsArbitrary}, are given in the full volume by
\begin{align}\label{eq:CvsGamma}
C_{vs}^\Gamma(t) &= \sum_{\mathbf{x}} \langle\, \pi_{vs}^\Gamma(\mathbf{x},t)\, 
		\pi_{vs}^{\Gamma \dagger}(\mathbf{0},0)\ \rangle
	\nonumber\\&
	= \sum_{n=0}^{\infty} \left[ A_n^\G +(-1)^{n_t} B_n^\G  \right] \textrm{cosh}\Big(m_{n} (t- N_t/2) \Big)\, .
\end{align}
In Appendix~\ref{app:Cvs}, we use the symmetries of the mixed domain-wall valence and staggered sea lattice action to show that the lightest mass state which dominates the long time behavior of these mixed meson correlation functions are always pions, such that for all $\Gamma$ considered in Eq.~\eqref{eq:GammaMvs},
\begin{equation}
	m_0 = \tilde{m}_{vs}\, .
\end{equation}
%
We find that the situation with Dirichlet boundary conditions in time is different.  In this case we find that the long Euclidean time behavior of the mixed meson correlation function is given by
\begin{align}\label{eq:CvsDBC}
C_{vs}^{\Gamma,D}(t) \simeq&\ A_{\pi_{vs}}^D e^{-\tilde{m}_{vs} t} 
	+ (-1)^{n_t} B_{\pi_{vs}}^D e^{-\tilde{M}_{vs}t}
\end{align}
with $\tilde{m}_{vs} \neq \tilde{M}_{vs}$.  For fermions with anti-periodic boundary conditions in time, there are two interpolating operators which couple to the mixed pions that are related by the time-doubling symmetry of the staggered action, in fact requiring $\tilde{m}_{vs} = \tilde{M}_{vs}$, as we observe with our full-volume calculations (see Appendix~\ref{app:Cvs} for details).  The Dirichlet boundary conditions break this time-doubling symmetry, which in position space, is the time-shift symmetry.  We hypothesize this symmetry breaking is responsible for the observed lifting of the degeneracy between straight and oscillating states.  In Appendix~\ref{app:MvsDBCs} we collect our results for $\tilde{m}_{vs}$ and $\tilde{M}_{vs}$, demonstrating that both states must be pions, as they are too light to be anything else.  Furthermore, one notices that the splitting in the square of these masses is independent of the pion mass, within errors, indicative that the splitting is due to a lattice spacing correction, consistent with expectations from MAEFT.  As an important test of systematics, on the $m_l=010$, $m_s=050$ coarse ensembles, we have performed our calculations using a set of full-volume propagators satisfying anti-periodic boundary conditions in time, as well as the half volume calculation with a Dirichlet wall.  In this case we find that
\begin{align}
	\tilde{m}_{\pi_{vs}} &= 0.283(3) \qquad \textrm{for time Dirichlet BCs}\, ,
	\nonumber\\
	\tilde{m}_{\pi_{vs}} &= 0.285(3) \qquad \textrm{for time anti-periodic BCs}\, .
\end{align}
Given that the extracted mixed meson mass is the same within errors, the issue of the boundary conditions is tangential to the focus of this work, and we do not address it further in the main text.  In Appendix~\ref{app:MvsDBCs} we provide further details of the connection between the breaking of the time-shift symmetry and the related change in staggered quark taste.

\section{Analysis and Results : Mixed meson mass renormalization\label{sec:results}}

To determine the meson masses, we use a time-correlated $\chi^2$-minimization fitting routine.  We perform both standard and jackknife fitting methods.  We fit the leading exponentials (cosh functions) for the domain-wall and staggered mesons to the two point correlation functions and the leading two exponentials for the mixed meson two-point correlation functions.  For the full volume mixed meson correlation functions, we use both a single cosh fit with a straight as well as oscillating amplitude and we also perform the fit allowing for two non-degenerate masses.  Explicitly, we fit the domain-wall pion correlation function with
\begin{align}
	A_\pi \textrm{cosh}\Big( m_\pi(t-N_t/2) \Big)\qquad& \textrm{ for PBCs}
	\nonumber\\
	A_\pi^D e^{-m_\pi t}\qquad& \textrm{ for DBCs}\, .
\end{align}
For the staggered Goldstone meson correlation function, we fit with an identical form,
\begin{align}
	A_{\pi_5} \textrm{cosh}\Big( m_{\pi_5}(t-N_t/2) \Big)\qquad& \textrm{ for PBCs}
	\nonumber\\
	A_{\pi_5}^D e^{-m_{\pi_5} t}\qquad& \textrm{ for DBCs}\, .
\end{align}
For the mixed meson correlation functions with DBC we fit the masses with the two exponentials
\begin{equation}
	A_{\pi_{vs}}^D e^{-\tilde{m}_{\pi}^{vs}t} +(-1)^{n_t} B_{\pi_{vs}}^D e^{-\tilde{M}_{\pi}^{vs}t}
\end{equation}
For the full volume lattices we fit to both
\begin{equation}\label{eq:CvsFitFunc}
	\Big[ A_{\pi_{vs}} +(-1)^{n_t} B_{\pi_{vs}} \Big] \textrm{cosh}\Big( \tilde{m}_{\pi}^{vs}(t-N_t/2) \Big)
\end{equation}
as well as 
\begin{equation}
	A_{\pi_{vs}} \textrm{cosh}\Big( \tilde{m}_{\pi}^{vs}(t-N_t/2) \Big)
	+(-1)^{n_t} B_{\pi_{vs}} \textrm{cosh}\Big( \tilde{M}_{\pi}^{vs}(t-N_t/2) \Big)
\end{equation}
to explicitly verify the degeneracy of the mixed pion masses, $\tilde{m}_{\pi}^{vs}$ and $\tilde{M}_{\pi}^{vs}$ with the full volume propagators.  The results of these fits are collected in Table~\ref{tab:results}.  The central values are determined by standard error analysis and the error bars by a jackknife analysis.%
\footnote{
In Appendix~\ref{app:pisfromrhos} we present the results of fitting the various mixed meson two-point correlation functions and in Appendix~\ref{app:MvsDBCs} we present the results of the mass splitting between the two light mixed pions with DBCs.} 
%
%
\begin{table}[t]
\caption{\label{tab:results} 
Pion masses and $\D (a m)^2 = (am_\pi^{vs})^2 -\frac{1}{2}\big[ (am_{\pi})^2 +(am_{\pi_5})^2 \big]$.  We present our results for the masses determined with the fitting routine described in Sec.~\ref{sec:results}.  We compute the additive mixed meson mass renormalization both with the staggered meson masses we determined for this work as well as by inputting the staggered meson masses from Ref.~\cite{Bernard:2001av}, see text for details.}
\begin{ruledtabular}
\begin{tabular}{cccccc}
Ensemble & pion & result & fit range & $\chi^2$ / d.o.f. & Q \\
\hline
m050 : DBC & $m_\pi^{vs}$ & 0.500(2) & 6--13 & 4.9 / 4 & 0.29 \\
m040 : DBC & $m_\pi^{vs}$ & 0.457(2) & 6--13 & 6.4 / 4 & 0.17 \\
m030 : DBC & $m_\pi^{vs}$ & 0.412(2) & 9--13 & 1.1 / 1 & 0.29 \\
m020 : DBC & $m_\pi^{vs}$ & 0.353(3) & 9--14 & 0.54 / 2 & 0.76  \\
m010 : DBC & $m_\pi^{vs}$ & 0.283(3) & 5--12 & 5.0 / 4 & 0.29 \\
\hline
m050 : DBC & $m_\pi$ & 0.477(1) & 6--15 & 2.9 / 8 & 0.94 \\
m040 : DBC & $m_\pi$ & 0.433(2) & 5--11 & 7.1 / 5 & 0.22  \\
m030 : DBC & $m_\pi$ & 0.375(1) & 7--14 & 7.8 / 6 & 0.25 \\
m020 : DBC & $m_\pi$ & 0.315(1) & 6--14 & 3.9 / 7 & 0.79  \\
m010 : DBC & $m_\pi$ & 0.224(1) & 6--15 & 6.1 / 8 & 0.64  \\
\hline
m050 : DBC & $m_{\pi_5}$ & 0.485(2) & 7--10 & 1.7 / 2 & 0.43 \\
m040 : DBC & $m_{\pi_5}$ & 0.440(2) & 5--8 & 1.2 / 2 & 0.55  \\
m030 : DBC & $m_{\pi_5}$ & 0.381(2) & 6--8 & 0.16 / 1 & 0.69 \\
m020 : DBC & $m_{\pi_5}$ & 0.319(2) & 4--6 & 0.22 / 1 & 0.64   \\
\hline
m010 : Full--V & $m_\pi^{vs}$ & 0.285(3) & 6--31& 20 / 23 & 0.65 \\
m010 : Full--V & $m_\pi$ & 0.2235(05) & 7--31 & 30 / 23 & 0.14 \\
m010 : Full--V & $m_{\pi_5}$ & 0.2248(03) & 13--32 & 17 / 18 & 0.51 \\
\end{tabular}
\begin{tabular}{ccccccc}
Ensemble & m010(Full--V) & m010 & m020 & m030 & m040 & m050 \\
\hline
$\D(am)^2$ & 0.031(1) & -- & 0.024(2) & 0.026(2) & 0.018(2) & 0.019(2) \\
$\D(am)^2_{stag.\ input}$ & 0.031(1) & 0.030(2) & 0.027(2) & 0.028(2) & 0.020(2) & 0.019(2)
\end{tabular}
\end{ruledtabular}
\end{table}

We find that the Dirichlet boundary conditions cause significant difficulties in extracting parameters from the staggered pion two-point correlation functions.%
\footnote{As a reminder, most of the domain-wall propagators we used came with Dirichlet boundary conditions.  We did not have the computing time to make our own, and so we borrowed them from the LHP and NPLQCD collaborations.  The choice of Dirichlet BCs for the the staggered propagators we did compute was one of convenience.  Clearly, this would not be a good choice for performing staggered lattice QCD calculations.} 
As can be seen from Table~\ref{tab:results}, we are only able to determine the staggered pion mass over a very short range of time as compared to the much greater time lengths used in typical staggered lattice computations with anti-periodic in time boundary conditions, for example in Refs.~\cite{Bernard:2001av,Aubin:2004fs}.  The difficulty arises from a combination of factors.  Because we are calculating pion two-point correlation functions, the backwards propagating states from the Dirichlet wall are light enough that they contribute to the signal to such an extent that they pollute the extraction of the forward propagating signal we are interested in even at short times from the source.  This problem is then amplified by the time-oscillatory nature of the staggered correlation functions, becoming more problematic for lighter quark masses.  We were not able to determine a mass of the staggered Goldstone pion (taste--5) on the m010 lattices.  On the m020--m050 lattices the staggered pion mass we determine is consistent with those in Ref.~\cite{Bernard:2001av} within our errors.  For both the mixed and domain-wall pion, we find that the masses we calculate on the DBC lattices are consistent with those determined on the full volume lattices, as is seen in Table~\ref{tab:results}.

%
%
\subsection{Determination of mixed meson parameters \label{sec:MvsAnalysis}}
To determine the mixed meson mass renormalization we perform a jackknife analysis of the difference
\begin{equation}\label{eq:DmsqLatt}
	\Delta(am)^2 = (a\tilde{m}_{\pi}^{vs})^2 -\frac{1}{2}(am_\pi)^2 -\frac{1}{2}(am_{\pi_5})^2\, .
\end{equation}
The results of this analysis are collected in Table~\ref{tab:results}.  
For the reasons discussed in the previous section, when we calculate this quantity we use the values of the staggered meson determined in Ref.~\cite{Bernard:2001av}.  In this case, the uncertainty in the input staggered pion masses is much less than the errors we determine for the mixed and domain-wall pions and therefore can be ignored.  We determine this additive mixed meson mass renormalization using both the DBC as well as the full volume propagators on the m010 lattices.  As is clear in both cases, 
there is an upwards trend of the mass renormalization with lighter quark mass, a signal of NLO effects, $\mc{O}(a^2 m_\pi^2)$.  To analyze these results we use several different fits.  First we fit the NLO MA extrapolation formula given in Eq.~\eqref{eq:DmvsNLO} with a $\chi^2$-minimization routine to the lightest three data points, the lightest four and all five data points given in Table~\ref{tab:results}, for both sets of m010 results.  In order to perform these fits, we use the values of the staggered taste splitting listed in Eq.~\eqref{eq:tasteSplit} which are taken from Ref.~\cite{Aubin:2004fs}.  We also input the values of the axial and vector hairpin couplings for which updated values can be found in Ref.~\cite{Bernard:2005ei}.  
These fits are all performed at the scale $a\mu=1$ and with the choice $af=0.103$.  The difference between using this value for $af$, the value in the chiral limit, or the value at another pion mass is higher order in the MAEFT than we are working.  Choosing a different renormalization scale will adjust the parameters of the MA fit but leave the mixed meson mass renormalization unchanged as this is a scale independent quantity.  As an example, the NLO MA fit to the lightest four points calculated with the propagators satisfying DBCs at the scale $a\mu=0.5$ gives $a^2(a^2\D_\mathrm{Mix}) = 0.0392(16)$ and $a^2(a^2L_{a^2m}) = -0.081(13)$ which directly verifies that the scale dependent contribution from the term $a^2(a^4 L_{a^4}(\mu))$, in Eq.~\eqref{eq:DmvsNLO}, is small compared to the LO term, $a^2\D_\mathrm{Mix}$.%
\footnote{The $a^2\D_\mathrm{Mix}$ term is scale independent.} 
This also demonstrates that the size of these unphysical counterterms are fairly insensitive to the renormalization scale.
We present the results of these fits in Table~\ref{tab:MAparams}.
%
%
\begin{table}[t]
\caption{\label{tab:MAparams} 
Determination of mixed action and quadratic fit parameters.  The upper table contains the fits in which we use the DBC propagators for the m010 lattices.  The lower table contains the fits in which we use the full volume propagators for the m010 lattices.  In both cases, we input the staggered pion mass values from Ref.~\cite{Bernard:2001av}.} 
\begin{ruledtabular}
\begin{tabular}{c|c|c}
\phantom{Full Volume} 
& Fits to NLO MA formula, Eq.~\eqref{eq:DmvsNLO} 
& Fits to quadratic formula, Eq.~\eqref{eq:quadFitFunc}
\end{tabular}
\begin{tabular}{c|ccc|ccc}
DBCs & $a^2(a^2\D_\mathrm{Mix})$ & $a^2(a^2L_{a^2m})$ & Q
& $a^2(a^2\d_\mathrm{Mix})$ & $a^2(a^2\ell_{a^2m})$ & Q \\
\hline
m010--m030 & 0.0383(15) & -0.054(31) & 0.30 & 0.0309(29) & -0.029(28) & 0.33 \\
m010--m040 & 0.0396(11) & -0.087(19) & 0.25 & 0.0336(22) & -0.064(17) & 0.19 \\
m010--m050 & 0.0394(10) & -0.084(13) & 0.42 & 0.0336(19) & -0.063(12) & 0.35 \\
\hline\hline 
Full Volume & & & & & &  \\
\hline
m010--m030 & 0.0394(10) & -0.074(25) & 0.19 & 0.0332(22) & -0.048(25) & 0.22 \\
m010--m040 & 0.0401(08)& -0.094(15) & 0.25 & 0.0350(17) & -0.072(14) & 0.20 \\
m010--m050 & 0.0399(07) & -0.090(11) & 0.40 & 0.0347(15) & -0.070(10) & 0.36 \\
\end{tabular}
\end{ruledtabular}
\end{table}
We additionally use a $\chi^2$-minimization to fit the same sets of data to a quadratic form,
\begin{equation}\label{eq:quadFitFunc}
	\D(am)^2 = a^2(a^2\d_\mathrm{Mix}) + (am_\pi)^2 a^2\ell_{a^2m}\, ,
\end{equation}
for which we also present the results of our analysis in Table~\ref{tab:MAparams}.  

\begin{figure}[!t]
\begin{tabular}{cc}
\includegraphics[width=0.47\textwidth]{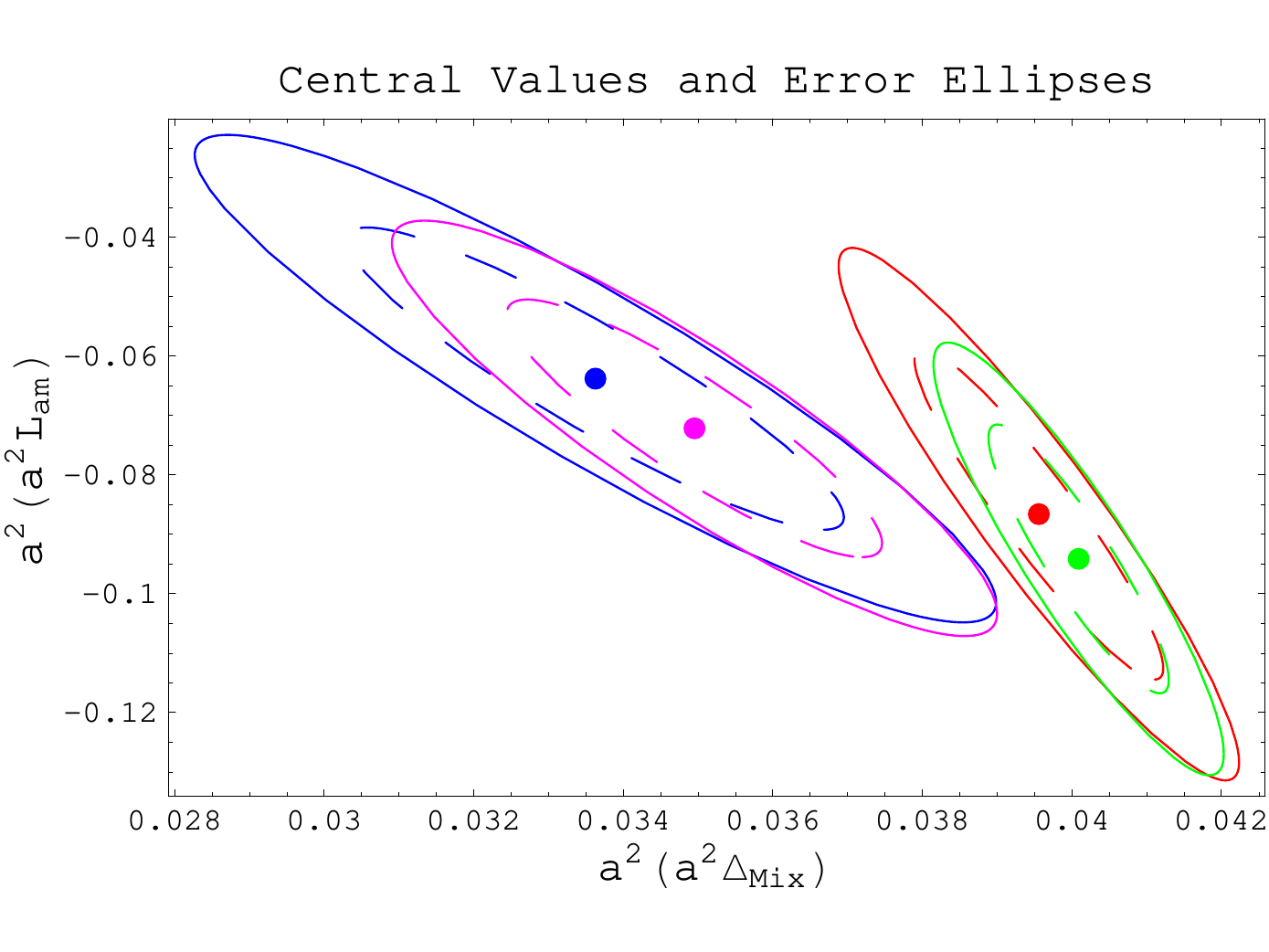}
&\includegraphics[width=0.485\textwidth]{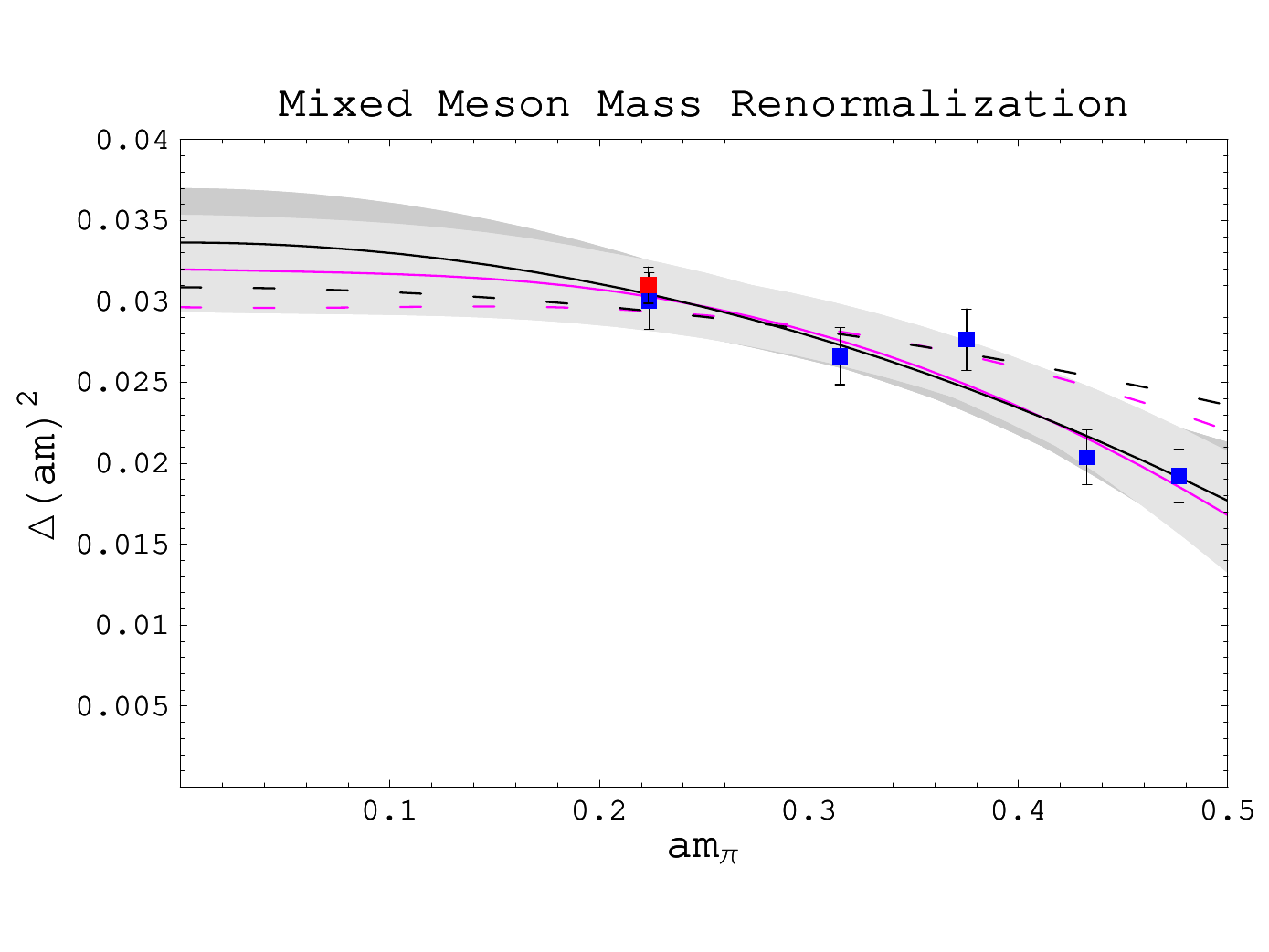}
\\
$(a)$ & $(b)$ 
\end{tabular}
\caption{\label{fig:asqDMix} (color online) 
This is a plot of the central result of our paper.  In Figure~\ref{fig:asqDMix}$(a)$ we present the central values, 68\% (dashed lines) and 95\% (solid lines) confidence intervals for four different fits, see Sec.~\ref{sec:MvsAnalysis} for details.  In Figure~\ref{fig:asqDMix}$(b)$, we plot both the results of our computation as well as best fit curves and 68\% confidence bands as a function of the domain-wall pion mass, $am_\pi$.  The darker curve and error band are the results of the quadratic fit to the lightest 4 data points and the lighter curve (and error band) represent the results of fitting to the NLO MA formula.  The dashed curves represent the same fits respectively but to the lightest three data points.}
\end{figure}

Fitting to these two different functions gives rise to different values of the unknown (counter)-terms, as is seen in Figure~\ref{fig:asqDMix}$(a)$, where we display the central values, 68\% (dashed ellipses) and 95\% (solid ellipses) confidence intervals for four different fits.  The error ellipses on the left of Fig.~\ref{fig:asqDMix}($a$) correspond to fitting the data in Table~\ref{tab:results} to the quadratic form, Eq.~\eqref{eq:quadFitFunc}.  The larger ellipse (blue) is fit to the lightest four data points including the m010 point calculated with the DBC propagators.  The smaller ellipse (purple) corresponds to a fit of the lightest four points but using the results of the full volume propagators for the m010 calculation.  The right two error ellipses correspond to fitting our results of the lightest four mass points to the NLO mixed action formula, Eq.~\eqref{eq:DmvsNLO} for the results obtained with DBC propagators (larger red) and with the full volume m010 propagators (smaller green) respectively.  As is seen by these error ellipses, there is no significant difference in using either the results calculated with the DBC propagators or the full volume propagators.
%
%

In Figure~\ref{fig:asqDMix}($b$) we plot the resulting mixed meson mass renormalization, Eq~\eqref{eq:DmsqLatt} and corresponding 68\% confidence error bands resulting from the quadratic and NLO mixed action fits to the lightest four data points with all results calculated with the DBC propagators, as well as the results of our calculation listed in Table~\ref{tab:results}.  For the m010 point ($am_\pi \simeq 0.2235$), the lower (blue) point is calculated with the DBC propagators and the slightly higher (red) point is calculated with the full volume propagators.  The darker (black) curve and error band correspond to the quadratic fit while the lighter (purple) curve and error band correspond to the NLO mixed action fit.  We additionally plot the resulting dashed curves for the same quadratic (black dashed curve) and NLO mixed action (lighter dashed curve) fits to the lightest three data points. 
As can be seen from Figure~\ref{fig:asqDMix}$(b)$, at the 68\% confidence level, these fits are all consistent (which also includes all fits listed in Table~\ref{tab:MAparams}).  Furthermore, for the lightest three data points they are equivalent, with very similar 68\% error bands.  In order to distinguish between these two curves one would need at least an order of magnitude increase in statistics.  
Therefore we conclude that within our precision, one can parameterize the mixed meson mass renormalization with either the NLO mixed action formula, Eq.~\eqref{eq:DmvsNLO}, or the quadratic formula, Eq.~\eqref{eq:quadFitFunc}: 
%
\footnote{We note that in Ref.~\cite{Bunton:2006va} an estimate of this quantity was made by comparing the MA extrapolation formula of the pion form-factor to a MA lattice simulation of this quantity, in which the mass renormalization was estimated to be $a^2\D_\mathrm{Mix} \sim (380 \textrm{ MeV})^2$.  In addition, an independent preliminary lattice calculation similar to this work is consistent with our results~\cite{RJC:AllHands2007}.} 
\begin{align}
	\D(am)^2 &= 0.040(1) - 0.09(2)(am_\pi)^2 + \chi_2^{NLO}\, ,\label{eq:MAfitResult}
	\\ &\quad \textrm{or} \nonumber\\
	\D(am)^2 &= 0.034(2) -0.06(2)(am_\pi)^2\, ,\label{eq:QUADfitResult}
\end{align}
where $\chi_2^{NLO}$ is slightly different from $\chi^{NLO}$ and can be determined with Eq.~\eqref{eq:DmvsNLO}.  In these expressions, we have chosen the fits to the lightest four points for which all the mass renormalizations were computed with the DBC propagators and the staggered pion masses were taken from Ref.~\cite{Bernard:2001av}.  We re-emphasize that these are not distinguishable to the resulting functions in which the full volume propagators are used in the calculation.  Using Eq.~\eqref{eq:CMix}, we can then determine $C_\mathrm{Mix}$ from Eq.~\eqref{eq:MAfitResult}, however, we emphasize that what is more important is the determination of the mass renormalization itself, Eq.~\eqref{eq:DmsqLatt}.  We find (recalling $a^2 \D_\mathrm{Mix} = a^2 16 C_\mathrm{Mix} / f^2$),
\begin{align}
\D_\mathrm{Mix} &\simeq (708 \textrm{ MeV})^4\, ,
\nonumber\\
\frac{C_\mathrm{Mix}}{f^2} &\simeq (354 \textrm{ MeV})^4\, ,
\end{align}
and choosing $f=132$~MeV,
\begin{equation}
C_\mathrm{Mix} \simeq (255 \textrm{ MeV})^6\, .
\end{equation}	
It will be interesting to compare these same parameters determined on the fine MILC lattices.

%
%
\section{Discussion \label{sec:concl}}
We have calculated the additive mass renormalization of mixed mesons composed of one domain-wall valence fermion and one staggered sea fermion for the asqtad-improved~\cite{Orginos:1998ue,Orginos:1999cr}, publicly available MILC lattices~\cite{Bernard:2001av} which are relevant to a large number of recent mixed action lattice calculations~\cite{Renner:2004ck,Bonnet:2004fr,Beane:2005rj,Edwards:2005kw,Edwards:2005ym,Beane:2006mx,Beane:2006pt,Beane:2006fk,Beane:2006kx,Alexandrou:2006mc,Beane:2006gj,Edwards:2006qx,Beane:2006gf}.  This is an important unphysical parameter, or lattice artifact to determine for a given mixed action lattice computation.  It modifies the mass of the mixed \textit{valence-sea} pions and therefore plays a central role in the low energy dynamics of all hadronic correlation functions, entering generally all extrapolation formula at the one-loop level determined with the appropriate mixed action effective field theory~\cite{Chen:2007ug}. 

The chiral extrapolations of the observables computed in this mixed action scheme~\cite{Renner:2004ck,Edwards:2005ym} have not included this mass renormalization as it has not been previously determined.  Now one can make use of either Eqs.~\eqref{eq:MAfitResult} or \eqref{eq:QUADfitResult} to determine the error in neglecting this mass renormalization in performing chiral extrapolations and  hopefully improve the uncertainty in many of the observables calculated to date~\cite{Renner:2004ck,Bonnet:2004fr,Beane:2005rj,Edwards:2005kw,Edwards:2005ym,Beane:2006mx,Beane:2006pt,Beane:2006fk,Beane:2006kx,Alexandrou:2006mc,Beane:2006gj,Edwards:2006qx,Beane:2006gf}.  In particular it is known that the pseudo scalar decay constants depend upon the mixed meson masses at next-to leading order~\cite{Bar:2005tu}.  Additionally, it is known that in mixed action lattice calculations, the $I=3/2\ K\pi$ scattering length~\cite{Chen:2006wf}, the pion form factor~\cite{Bunton:2006va}, baryon masses~\cite{Tiburzi:2005is}, nucleon axial charge~\cite{Jiang:2007sn}, nucleon twist-2 matrix elements, the nucleon to delta electromagnetic transitions and the neutron electric dipole moment~\cite{Chen:2007ug} to name a few, all depend upon the additive mixed meson mass renormalization as they depend upon the mixed meson masses at the one loop level in the mixed action effective field theory.  Therefore, with this additive mixed meson mass renormalization known, one can perform mixed action chiral extrapolations of all these quantities, as well as any others computed with domain-wall valence fermions and the asqtad-improved staggered sea fermions at this lattice spacing, $a\sim0.125$ fm.  Clearly, Eq.~\eqref{eq:QUADfitResult} is the most useful for including this systematic effect in these extrapolations.  To be used in the type of mixed action calculations this work is addressing~\cite{Renner:2004ck,Bonnet:2004fr,Beane:2005rj,Edwards:2005kw,Edwards:2005ym,Beane:2006mx,Beane:2006pt,Beane:2006fk,Beane:2006kx,Alexandrou:2006mc,Beane:2006gj,Edwards:2006qx,Beane:2006gf}, one takes a given mixed action extrapolation formulae~\cite{Chen:2007ug}, and wherever there is a mixed valence-sea meson in the expression, one makes the replacement (in lattice units)
\begin{align}
	\tilde{m}_{\pi_{vs}}^2 &\rightarrow m_\pi^2 + 0.034(2) - 0.06(2) m_\pi^2\, ,
	\nonumber\\
	\tilde{m}_{K_{vs}}^2 & \rightarrow m_K^2 + 0.034(2) - 0.06(2) m_\pi^2\, ,
\end{align}
where $m_\pi$ and $m_K$ are the corresponding pion and kaon masses calculated from domain-wall correlation functions.

%
%
\begin{acknowledgments}
The computations for this work were performed at The College of William and Mary on the \textit{Cyclades Cluster} purchased with funds from the Jeffress Memorial Trust Grant No. J-813.  AWL would like to thank the physics department at the University of Washington for allowing him to use their \textit{Grad Student Cluster} where code for this project was tested.  AWL would also like to thank Paulo Bedaque for many useful conversations, Maarten Golterman for useful correspondence and Claude Bernard for providing a program to convert the staggered taste splitting and hairpin couplings into lattice units.  The work of KO was supported in part by the U.S.~Dept.~of Energy contract No.~DE-AC05-06OR23177 (JSA) and contract No.~DE-FG02-04ER41302.  The work of AWL was supported under DOE grant DE-FG02-93ER-40762.
\end{acknowledgments}

%
%
\appendix

%
%
\section{Mixed Meson Two-Point Correlation Functions\label{app:Cvs}}

Understanding the two-point correlation functions of mixed mesons composed of one ``Wilsonized"-staggered fermion and one domain-wall fermion, whose interpolating operators are given in Eq.~\eqref{eq:pivsArbitrary},
requires an understanding of the symmetries of the staggered lattice action.  This has not been previously worked out before, although pieces of the required formalism can be found for example in Refs.~\cite{Kawamoto:1981hw,Sharatchandra:1981si,Kluberg-Stern:1983dg,vandenDoel:1983mf,Golterman:1984cy,Wingate:2002fh}.

In the above interpolating field, $\psi_{dw}(x)$ is a domain-wall fermion operator and $\psi_s(x)$ is a ``Wilsonized" staggered fermion operator as discussed in the text in Sec.~\ref{sec:LattCorrelators}, and $\Gamma$ is one of the set of gamma matrices intended to provide the spin and parity quantum numbers of a given meson.  As we shall demonstrate, because of the peculiarities of the staggered lattice action, for all $\Gamma$ in 
\begin{equation}
	\Gamma = \{ \mathbb{I}, \g_5\, ,\,  \g_\mu\, ,\,  \g_5 \g_\mu\, ,\,  \g_\mu \g_\nu \}\,
\end{equation}
the corresponding mixed meson interpolating operators have a non-vanishing coupling to pions.  In particular we will show that Eqs.~\eqref{eq:CvsGamma} provides the correct description of the long time behavior of the mixed two-point correlation functions.

We begin with a review of the doubling symmetries of the naive fermion action, which are also symmetries of the staggered action and provides a basis for the following discussion.  Much of the first part of this discussion can be found in Ref.~\cite{Wingate:2002fh}.

\bigskip
The free, naive and staggered fermion actions are invariant under the following ``doubling" transformations
\begin{align}
	\psi(x) \longrightarrow&\ e^{i \pi_g \cdot x}\, M_g\, \psi(x)\, ,
	\nonumber\\
	\bar{\psi}(x) \longrightarrow&\ \bar{\psi}(x)\, e^{i \pi_g \cdot x}\, M^\dagger_g\, ,
	\nonumber\\
	S \longrightarrow S =&\ a^4 \sum_{x} \bar{\psi}(x)\, \Big[ \frac{\nabla_\mu}{a} \gamma_\mu 
		+m \Big]\, \psi(x)
\end{align}
where in four space-time dimensions $g$ are the 16 elements of the set of ordered lists with up to 4 elements (including the empty set):
\begin{equation}
	G = \{ g: g = ( \mu_1, \mu_2, \mu_3, \mu_4),\, \mu_1 < \mu_2 < \mu_3 < \mu_4 \}\, ,
\end{equation}
\begin{equation}\label{eq:Mg}
	M_g = \prod_{\mu \in g} M_\mu\qquad, \qquad \textrm{ with } M_\mu = i \g_5 \g_\mu\, .
\end{equation}
and the momentum shifts are given by
\begin{equation}\label{eq:pi_g}
	(\pi_g)_\mu = \left\{ \begin{array}{cc}
		\frac{\pi}{a}, & \mu \in g \\ \\
		0, & \mu \notin g 
		\end{array} \right. \, .
\end{equation}
Here we are using Hermitian Euclidean gamma matrices satisfying the anti-commutation relations, Eq.~\eqref{eq:EuclidCommute}.  To realize the doubling symmetry, one introduces 16 momentum space spinors,
\begin{align}\label{eq:naiveTastes}
	q^g(k) &= M_g \psi(k+\pi_g)\, ,
	\nonumber\\
	\bar{q}^g(k) &= \bar{\psi}(k+\pi_g) M^\dagger_g\, ,
\end{align}
which when combined with the periodicity of the integrand over the Brillouin zone, allows one to show that the naive action has a 16-fold degeneracy (in 4 space-time dimensions),
\begin{align}
S &= \int_{-\pi/a}^{\pi/a} \frac{d^4 k}{(2\pi)^4} \bar{\psi}(k) \left[ \sum_\mu i \g_\mu \frac{1}{a} 
	\textrm{sin}(k_\mu a) +m \right] \psi(k)
\nonumber\\&
	= \sum_{g\in G} \int_{-\pi/2a}^{\pi/2a} \frac{d^4 k}{(2\pi)^4} \bar{\psi}(k+\pi_g) \left[ \sum_\mu i \g_\mu 
		\frac{1}{a} \textrm{sin}([k+\pi_g]_\mu a) +m \right] \psi(k+\pi_g)
\nonumber\\&
	= \sum_{g\in G} \int_{-\pi/2a}^{\pi/2a} \frac{d^4 k}{(2\pi)^4} \bar{q}^g(k) \left[ \sum_\mu i \g_\mu 
		\frac{1}{a} \textrm{sin}(k_\mu a) +m \right] q^g(k)\, .
\end{align}

\bigskip
To connect the staggered and \textit{Wilsonized}-staggered fermions with the naive fermions, it is useful to use the Kawamoto-Smit transformation of the naive fermion action in position space~\cite{Kawamoto:1981hw}.  First one introduces new fields,
\begin{align}
	\Phi(x) &= \Omega^\dagger(x) \psi(x)\, ,
	\\
	\bar{\Phi}(x) &= \bar{\psi}(x) \Omega(x)\, ,
\end{align}
where
\begin{equation}
	\Omega(x) = \prod_{\mu=1}^{4} (\g_\mu)^{x_\mu / a}\, .
\end{equation}
The action expressed in terms of these new fields is then
\begin{equation}
	S_\Phi = a^4 \sum_{x,\mu} \sum_\alpha \bar{\Phi}_\a(x)\, \Big[ \eta_\mu(x)\, \frac{\nabla_\mu}{a}  
		+m \Big]\, \Phi_\a(x)\, ,
\end{equation}
with
\begin{equation}
	\eta_\mu(x) = (-)^{(x_1 +\dots + x_{\mu-1})/a}\, .
\end{equation}
In this form, the action is a sum over the four independent spin components of the Dirac fermion, $\Phi$, and one can immediately relate the single component staggered fermions to the \textit{Wilsonized}-staggered fermions or the naive fermions by making use of a constant spinor,  
\begin{equation}
	\Phi_\a(x) = C_\a\, \chi(x)\, .
\end{equation}
This also allows us to relate the propagators of the staggered and \textit{Wilsonized}-staggered fermions which is necessary to understand the mixed meson two-point correlation functions.  One finds the propagator from $x$ to $y$,
\begin{align}\label{eq:StagProp}
	D_{\psi_s}(x;y) &= \Omega(y)\, \mathbb{I}_{4}\, D_\chi(x;y)\, \Omega^\dagger(x)
	\nonumber\\&
	= \Omega(y)\, \Omega^\dagger(x)\, D_\chi(x;y)\, .
\end{align}
This analysis also allows one to easily derive the behavior of staggered meson two-point correlators by converting the various meson sources from the naive fermion form to their corresponding staggered form as is found in Ref.~\cite{Golterman:1985dz}.  We now describe the source of time-oscillatory behavior.

%
%
\subsection{Time oscillations from doubling symmetry}\label{sec:doublingSymm}
For our analysis of the mixed correlators, it will be convenient to examine the fermion fields in momentum space with the momentum split up amongst the different corners of the Brillouin zone.  To do this, we first introduce the subset of $G$ which contains the spatial momentum shifts only, $G_s$.  Using the periodicity of the momentum region over the Brillouin zone and Eq.~\eqref{eq:naiveTastes}, one has
\begin{align}\label{eq:stagMomentum}
\psi_s(x) &= \sum_{g_s \in G_s} e^{i \vec{\pi}_{g_s} \cdot x} \int_{-\pi/2a}^{\pi/2a} \frac{d^3 k}{(2\pi)^3} 
		e^{i \vec{k} \cdot x} M^\dagger_{g_s} q^{g_s}(k,t)\, ,
\end{align}
and similarly for the domain-wall fermion
\begin{equation}\label{eq:dwMomentumField}
\bar{\psi}_{dw}(x) = \sum_{g_s \in G_s} e^{-i \vec{\pi}_{g_s} \cdot x} \int_{-\pi/2a}^{\pi/2a} \frac{d^3 k}{(2\pi)^3} 
	e^{-i \vec{k}\cdot x} \bar{\psi}_{dw}(\vec{k}+\vec{\pi}_{g_s}, t)\, .
\end{equation}
The full doubling group is recovered with the product of the spatial doublers and the time-doublers, $g_s$ and $g_s g_t$, with $\pi_{g_t} = \frac{\pi}{a}(\mathbf{0},1)$ and 
\begin{equation}\label{eq:Mgt}
	M_{g_t} = i \gamma_5 \gamma_4\, .
\end{equation}  
However, to understand the source of time-oscillations present in the mixed meson correlators, it is useful to treat the time doublers separately, which we now do.  Using the time doubling symmetry, $g_t$, one can show that the staggered (and naive) field can be expressed as
\begin{align}\label{eq:stagTimeOsc}
\psi_s( \vec{k}+\vec{\pi}_{g_s}, t) &= \int_{-\pi/a}^{\pi/a} \frac{dk_4}{2\pi} e^{ik_4 t} 
	\psi(\vec{k}+\vec{\pi}_{g_s}, k_4)
\nonumber\\&
	= \int_{-\pi/2a}^{\pi/2a} \frac{dk_4}{2\pi} e^{ik_4 t} \Big[ \psi(\vec{k}+\vec{\pi}_{g_s}, k_4)
		+ e^{i\pi_{g_t} \cdot x} \psi(\vec{k}+\vec{\pi}_{g_s}, k_4 + \pi_{g_t}) \Big]
\nonumber\\&
	= \int_{-\pi/2a}^{\pi/2a} \frac{dk_4}{2\pi} e^{ik_4 t} \Big[ M^\dagger_{g_s} q^g(\vec{k}, k_4)
		+ (-1)^{n_t}\,  M^\dagger_{g g_t} q^{g g_t}(k, k_4) \Big]\, .
\end{align}
One can immediately see that this field contains both ``straight" as well as time-oscillating components.  
Close to the continuum limit, the time-oscillating quark field is of a different staggered taste than the straight state.  At finite lattice spacing, there are taste-changing interactions and so one can not create a source with definite taste.  However, we argue that at finite lattice spacing, the oscillating states will still arise from this particular taste of staggered fermion, or more precisely with a particular change in taste, with the momentum shifted to the ``time" corner of the Brillouin zone.  We will address these issues further when we analyze the two-point correlation functions with Dirichlet boundary conditions.

%
%
\subsection{Staggered symmetry and mixed meson correlation functions \label{app:doublingSTAG}}
To understand the behavior of the mixed meson two-point correlation functions, we must go beyond this, which requires a more detailed analysis of the staggered symmetries.  Constructing the staggered fermion action from the naive one amounts to a maximal diagonalization of the doubling symmetry group, $G$, in which one chooses a maximal subgroup $H \subset G$, for which the matrices, $M_h$, defined in Eq.~\eqref{eq:Mg}, are all commuting~\cite{Sharatchandra:1981si},
\begin{equation}
	[ M_h\, ,\, M_{h^\prime}] = 0\quad,\quad \forall\, \{h,h^\prime\} \in H\, .
\end{equation}
This provides constraining equations amongst the 16 fermion doublers, or ``tastes", reducing the fermion degrees of freedom, and leading to the relations
\begin{equation}\label{eq:stagFermSymm}
\psi_s(x) = e^{i\pi_h\cdot x} \hat{M}_h \psi_s(x)\, ,
\end{equation}
where
\begin{equation}
	\hat{M}_h = \frac{M_h}{\l_h}\, .
\end{equation}
To provide specific examples, we will use the set
\begin{equation}
	H = \{ g_h: g_h = \emptyset, (3), (1,2), (1,2,3) \}\, ,
\end{equation}
with
\begin{equation}
	\l_0 = \l_{12} = 1\quad,\quad \l_3=\l_{123} = i\, ,
\end{equation}
and therefore
\begin{equation}
	\hat{M}_\emptyset = \mathbb{I}_{4\times4} \quad,\quad
	\hat{M}_3 = i\g_5 \g_3 \quad,\quad
	\hat{M}_{12} = -i \g_1 \g_2 \quad,\quad
	\hat{M}_{123} = \g_4\, .
\end{equation}
To understand how the staggered symmetry modifies the analysis, we must return to the decomposition of the staggered field in momentum space, Eq.~\eqref{eq:stagMomentum}.  First, we learn that a time-momentum space equivalent of the the staggered symmetry, Eq.~\eqref{eq:stagFermSymm} is
\begin{equation}
	\psi_s(k,t) = \hat{M}_h \psi_s(k+\pi_h,t)\, .
\end{equation}
Reexamining the staggered field in momentum space, we find
\begin{align}
\psi_s(x) &= \int_{-\pi/2a}^{\pi/2a} \frac{d^3 k}{(2\pi)^3} 
	\bigg[ \sum_{h \in H} 
		e^{i (\vec{k}+\pi_{h}) \cdot x} \psi_s(\vec{k}+\pi_{h}, t)
		+\sum_{g_s \in G_s/H} e^{i (\vec{k}+\vec{\pi}_{g_s}) \cdot x} \psi_s(\vec{k}+\vec{\pi}_{g_s}, t)
		\bigg]
	\nonumber\\&
	=\int_{-\pi/2a}^{\pi/2a} \frac{d^3 k}{(2\pi)^3} 
	\bigg[ \sum_{h \in H} 
		e^{i \vec{k} \cdot x} \hat{M}_h^\dagger \psi_s(\vec{k}, t)
		+\sum_{g_s \in G_s/H} e^{i (\vec{k}+\vec{\pi}_{g_s}) \cdot x} \psi_s(\vec{k}+\vec{\pi}_{g_s}, t)
		\bigg] \label{eq:psi_s_h}
\end{align}
We then insert this field into the mixed meson interpolating operator, project onto zero spatial momentum, and use Eqs.~\eqref{eq:stagMomentum} and \eqref{eq:dwMomentumField} to find
\begin{align}
\pi_{vs}^\Gamma(t, \vec{p}=0) =&\ a^3 \sum_{\vec{x}} \bar{\psi}_{dw}(x) \Gamma \psi_s(x) 
\nonumber\\ 
	=&\  \sum_{h\in H} \int_{-\pi/2a}^{\pi/2a} \frac{d^3 k}{(2\pi)^3} 
		\bar{\psi}_{dw}(\vec{k},t)\, \Gamma\, \hat{M}_h^\dagger\, \psi_s(\vec{k}, t)
	\nonumber\\& 
	+\sum_{g_s\neq h, g_s \in G_s} \int_{-\pi/2a}^{\pi/2a} \frac{d^3 k}{(2\pi)^3} 
		\bar{\psi}_{dw}(\vec{k}+\vec{\pi}_{g_s},t) \Gamma \psi_s(\vec{k}+\vec{\pi}_{g_s},t)\, .
\end{align}
A key observation to make is that the domain-wall fermions do not share the time-doubling symmetry that the naive and staggered fermions have.  Therefore, the modes in the sum over $g_s \neq h$ represent mesons with momentum on the order of the cutoff, $p \sim \pi/a$ and can therefore be integrated out of the theory and neglected in the analysis.  The general mixed meson interpolating field couples to the following sources,
\begin{align}\label{eq:MvsG}
\pi_{vs}^\G(t) &=\int_{-\pi/2a}^{\pi/2a} \frac{d^3 k}{(2\pi)^3} \bigg[
		\bar{\psi}_{dw}(\vec{k},t)\, \G\, \psi_s(\vec{k}, t)
		+\bar{\psi}_{dw}(\vec{k},t)\, \G \gamma_4\, \psi_s(\vec{k}, t)
	\nonumber\\&\qquad\qquad\qquad\quad
		+i\bar{\psi}_{dw}(\vec{k},t)\, \G \g_5 \gamma_3\, \psi_s(\vec{k}, t)
		+i\bar{\psi}_{dw}(\vec{k},t)\, \G \gamma_2 \gamma_1\, \psi_s(\vec{k}, t)
		\bigg]\, .
\end{align}
To understand the corresponding states this excites, we consider the general mixed meson two-point correlation function,
\begin{align}\label{eq:CvsG}
C_{vs}^\Gamma(t,\mathbf{p}=0) &= a^3 \sum_{\mathbf{x}} 
	\langle\, \bar{\psi}_s(t,\mathbf{x})\, \Gamma^\dagger\, \psi_{dw}(t,\mathbf{x})\,
		\bar{\psi}_{dw}(0,\mathbf{0})\, \Gamma\, \psi_s(0,\mathbf{0})\, \rangle
	\nonumber\\& 
	= a^3 \sum_{\mathbf{x}} 
	\langle\, D_{dw}(0,\mathbf{0};t,\mathbf{x})\, \Gamma\, D_s(t,\mathbf{x};0,\mathbf{0})\, 
		 \Gamma^\dagger \, \rangle
	\nonumber\\& 
	= a^3 \sum_{\mathbf{x}} 
	\langle\, D_{dw}(0,\mathbf{0};t,\mathbf{x})\, \Gamma\, D_\chi(t,\mathbf{x};0,\mathbf{0})\, 
	\Omega^\dagger(t,\mathbf{x})\,  \Gamma^\dagger \, \rangle
	\nonumber\\& 
	= a^3 \sum_{\mathbf{x}} \e(x)
	\langle\, D_{dw}(0,\mathbf{0};t,\mathbf{x})\, D_\chi^*(0,\mathbf{0};t,\mathbf{x})\, 
	\Gamma\, \Omega^\dagger(t,\mathbf{x})\,  \Gamma^\dagger \, \rangle\, .
\end{align}
with
\begin{equation}
	\e(x) = (-1)^{(x_1 +x_2 +x_3 +t)/a}\, ,
\end{equation}
and $\langle\ \rangle$ denoting a sum over the color and spin indices and an average over gauge configurations.  Strictly speaking, this equation, and similar ones that follow, are only true in the limit of an infinite number of configurations.  The spin factor, 
$\Gamma\, \Omega^\dagger(t,\mathbf{x}) \Gamma^\dagger$ for the various 
$\Gamma$ will determine which staggered phases and quantum numbers a given source excites, for which the following relations are useful
\begin{align}
	\g_5\, \Omega^\dagger(t,\mathbf{x})\, \g_5 &= \e(x)\, \Omega^\dagger(t,\mathbf{x})
	\nonumber\\
	\g_\mu\, \Omega^\dagger(t,\mathbf{x})\, \g_\mu &= \eta_\mu(x)\, \zeta_\mu(x)\, 
		\Omega^\dagger(t,\mathbf{x})
\end{align}
with the remaining staggered phase factors given by
\begin{align}
	&\eta_1(x) = 1,& 
	&\eta_2(x) = (-1)^{x_1},& 
	&\eta_3(x) = (-1)^{x_1+x_2},& 
	&\eta_4(x) = (-1)^{x_1 +x_2 +x_3},&
	\nonumber\\
	&\zeta_1(x) = (-1)^{x_2 +x_3 +x_4},& 
	&\zeta_2(x) = (-1)^{x_3 +x_4},& 
	&\zeta_3(x) = (-1)^{x_4},& 
	&\zeta_4(x) = 1.&
\end{align}
In addition to determining the states excited by the sources in Eq.~\eqref{eq:MvsG}, we can also verify that the time-oscillating partners related by a change in the staggered quark taste, from Eq.~\eqref{eq:stagTimeOsc},%
\footnote{This can be equivalently understood working in position space and applying the time-doubling transformation, $\psi_s(x) \rightarrow i e^{i \pi t/a} \g_5\g_4\, \psi_s(x)$ and determining the change in the two-point correlation functions.} 
otherwise have the same quantum numbers, for which the Dirac structure is modified by $\g_4\g_5$, Eq.~\eqref{eq:Mgt}.  We find
\begin{subequations}
\begin{align}
C_{vs}^{\g_5}(t) &= a^3 \sum_{\mathbf{x}} \langle\, D_{dw}(0,\mathbf{0};t,\mathbf{x})\, 
	D_\chi^*(0,\mathbf{0};t,\mathbf{x})\, \Omega^\dagger(t,\mathbf{x})\, \rangle\, ,\label{eq:Cvsg5}
\\
C_{vs}^{\g_4}(t) &= (-1)^{t+1} a^3 \sum_{\mathbf{x}} \langle\, D_{dw}(0,\mathbf{0};t,\mathbf{x})\, 
	D_\chi^*(0,\mathbf{0};t,\mathbf{x})\, \Omega^\dagger(t,\mathbf{x})\, \rangle\, ,
\end{align}
\end{subequations}
\begin{subequations}
\begin{align}
C_{vs}^{\g_5\g_4}(t) &= a^3 \sum_{\mathbf{x}} \eta_4(x) \langle\, D_{dw}(0,\mathbf{0};t,\mathbf{x})\, 
	D_\chi^*(0,\mathbf{0};t,\mathbf{x})\, \Omega^\dagger(t,\mathbf{x})\, \rangle\, ,
\\
C_{vs}^{1}(t) &= (-1)^{t+1} a^3 \sum_{\mathbf{x}} \eta_4(x) \langle\, D_{dw}(0,\mathbf{0};t,\mathbf{x})\, 
	D_\chi^*(0,\mathbf{0};t,\mathbf{x})\, \Omega^\dagger(t,\mathbf{x})\, \rangle\, ,
\end{align}
\end{subequations}
\begin{subequations}
\begin{align}
C_{vs}^{\g_i}(t) &= a^3 \sum_{\mathbf{x}} (-1)^{x_i} \langle\, D_{dw}(0,\mathbf{0};t,\mathbf{x})\, 
	D_\chi^*(0,\mathbf{0};t,\mathbf{x})\, \Omega^\dagger(t,\mathbf{x})\, \rangle\, ,
\\
C_{vs}^{\g_i\g_j}(t) &= (-1)^{t+1} a^3 \sum_{\mathbf{x}} (-1)^{x_i+x_j} \eta_4(x) 
	\langle\, D_{dw}(0,\mathbf{0};t,\mathbf{x})\, D_\chi^*(0,\mathbf{0};t,\mathbf{x})\, 
	\Omega^\dagger(t,\mathbf{x})\, \rangle\, ,
\end{align}
\end{subequations}
\begin{subequations}
\begin{align}
C_{vs}^{\g_4\g_i}(t) &= a^3 \sum_{\mathbf{x}} (-1)^{x_i} \eta_4(x) \langle\, 
	D_{dw}(0,\mathbf{0};t,\mathbf{x})\, D_\chi^*(0,\mathbf{0};t,\mathbf{x})\, 
	\Omega^\dagger(t,\mathbf{x})\, \rangle\, ,\label{eq:Cvsg4gi}
\\
C_{vs}^{\g_5\g_i}(t) &= (-1)^{t+1} a^3 \sum_{\mathbf{x}} (-1)^{x_i} \eta_4(x) \langle\, 
	D_{dw}(0,\mathbf{0};t,\mathbf{x})\, D_\chi^*(0,\mathbf{0};t,\mathbf{x})\, 
	\Omega^\dagger(t,\mathbf{x})\, \rangle\, ,\label{eq:Cvsg5gi}
\end{align}
\end{subequations}
explicitly verifying both the time-oscillatory nature of the correlation functions related by a time-doubling transformation.

%
%
\subsection{Pions from $\rho$'s, $a_0$'s and other meson sources \label{app:pisfromrhos}}
%
%
\begin{figure}[t]
\begin{tabular}{cc}
\includegraphics[width=0.48\textwidth]{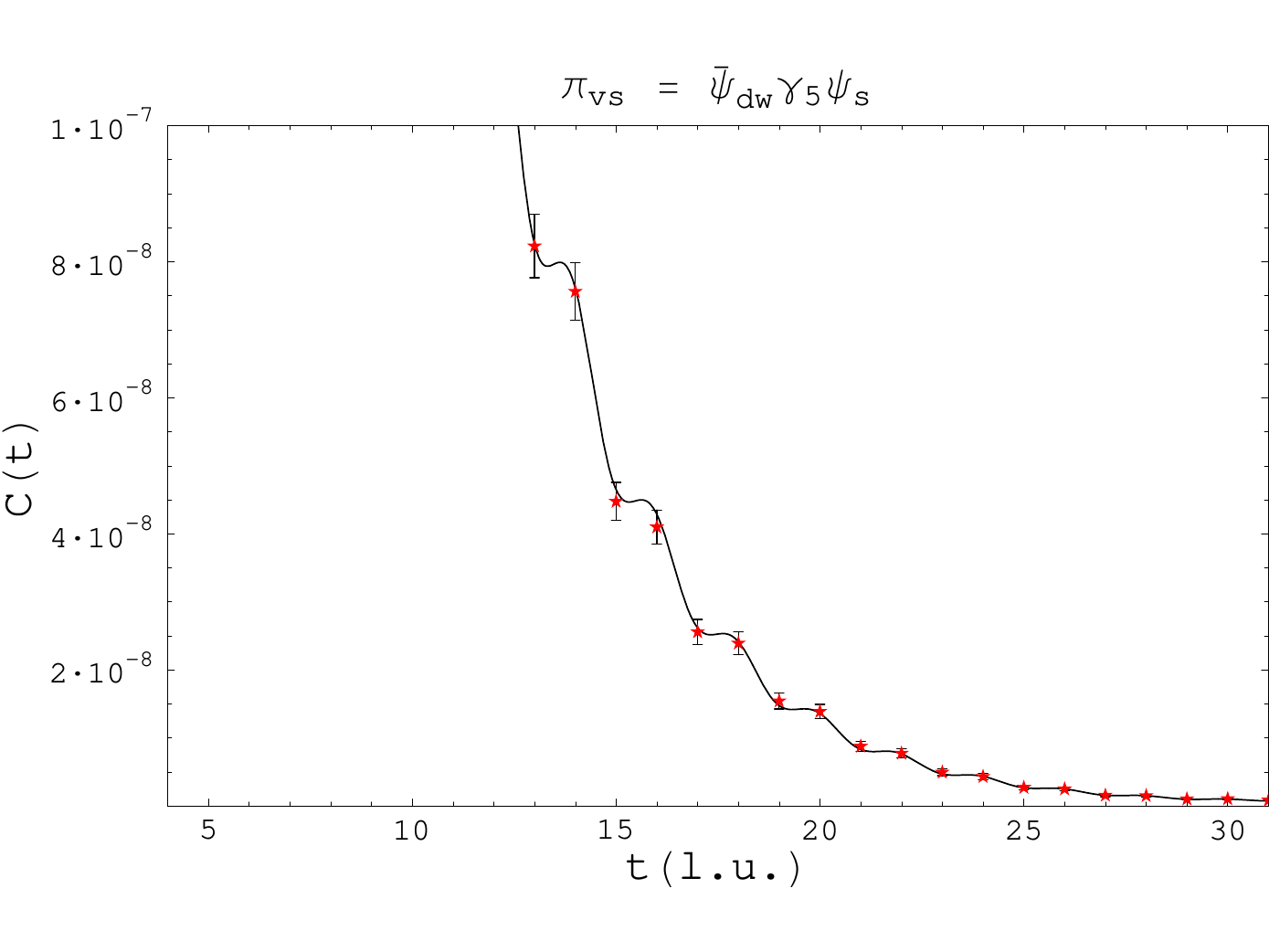}
&\includegraphics[width=0.48\textwidth]{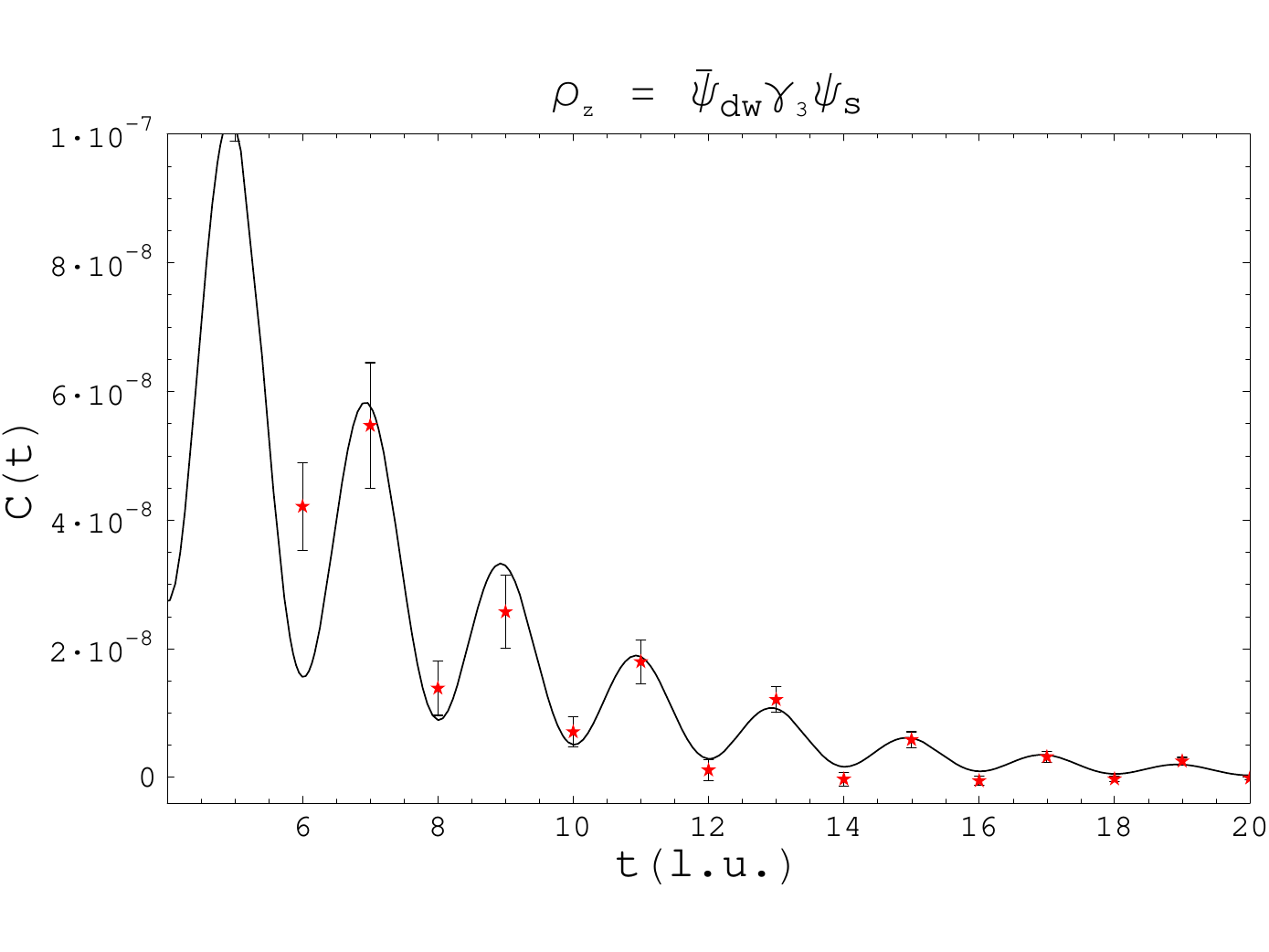}\\
(a) & (b)
\end{tabular}
\caption{\label{fig:MvsPionCorr} Correlation functions from mixed meson sources.  In Figure~\ref{fig:MvsPionCorr}$(a)$, we plot the correlator for the pseudo scalar source, $\pi_{vs} = \bar{\psi}_{dw} \g_5 \psi_s$.  In Figure~\ref{fig:MvsPionCorr}$(b)$ we plot the mixed ``rho" meson correlator, $\rho_z= \bar{\psi}_{dw} \g_3 \psi_s$.  The long time behavior of both correlators is dominated by single pion states, as demonstrated by the best fits to both with Eq.~\eqref{eq:CvsFitFunc}.  See Sec.~\ref{app:Cvs} for details and Table~\ref{tab:MIXEDresults} for results.}
\end{figure}

We can now show that pions couple to all of the naive mixed meson sources with Dirac structure given by Eq.~\eqref{eq:GammaMvs}, by inserting $\G = \{\mathbb{I}, \g_5, \g_3, \g_5\g_3, \g_1\g_2 \}$ into Eq.~\eqref{eq:MvsG}, and using Eq.~\eqref{eq:stagTimeOsc}, one can verify that all of the $\pi_{vs}^\G(x)$ interpolating fields couple to all the sources given in Eqs.~\eqref{eq:Cvsg5}--\eqref{eq:Cvsg5gi}.  
Because of the mixing of spin and taste degrees of freedom in the staggered action, computing the two-point correlation function given in Eq.~\eqref{eq:Cvsg5}, one is not only exciting states with pseudo-scalar quantum numbers.  Using the symmetries of the staggered action, we have shown that the source for this correlator couples not just to Eq.~\eqref{eq:Cvsg5}, but that Eq.~\eqref{eq:Cvsg5} is just one component of a larger matrix of correlation functions excited by the naive pion source, for which Eqs.~\eqref{eq:Cvsg5}--\eqref{eq:Cvsg5gi} are diagonal entries.  The naive pseudo-scalar source will have an overlap with the propagating pion, as one term of this source has the quantum numbers of an odd parity scalar, but likewise from Eq.~\eqref{eq:MvsG}, so will all the naive mixed meson sources.  Therefore we have shown that Eq.~\eqref{eq:CvsGamma} provides the correct description of the long-time behavior of all the mixed meson interpolating fields created with (anti) periodic boundary conditions considered in this work (anti periodic boundary conditions for the fermions and periodic for the mesons).  In principle, one can determine what linear combination of sources, $\pi_{vs}^\G$ will have an overlap with only the pion and other mesons by studying the matrix of correlation functions, disentangling the eigen-states excited by these sources.  This however, is beyond the scope of this work. 

\begin{figure}[t]
\begin{tabular}{cc}
\includegraphics[width=0.48\textwidth]{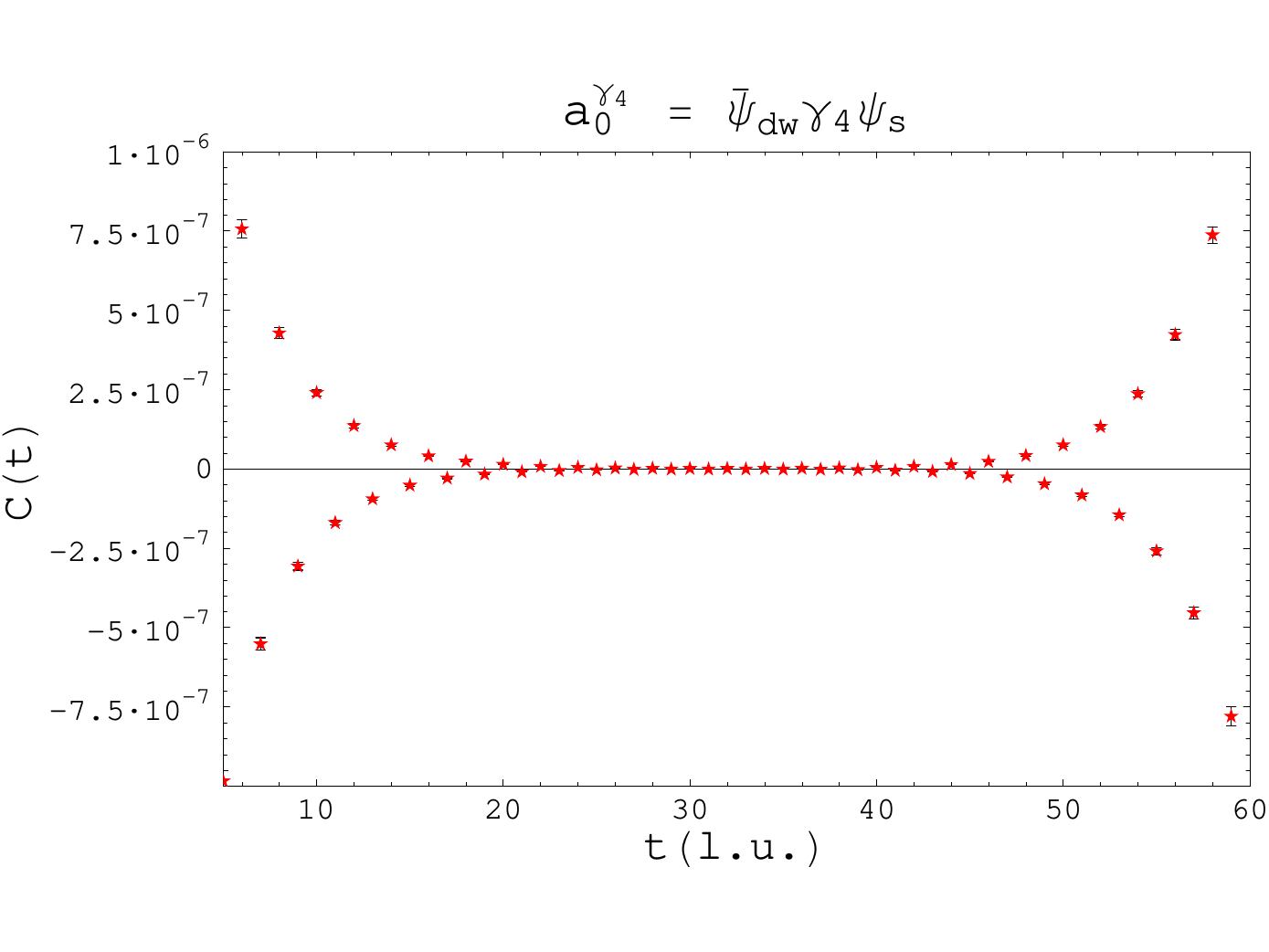}
&\includegraphics[width=0.48\textwidth]{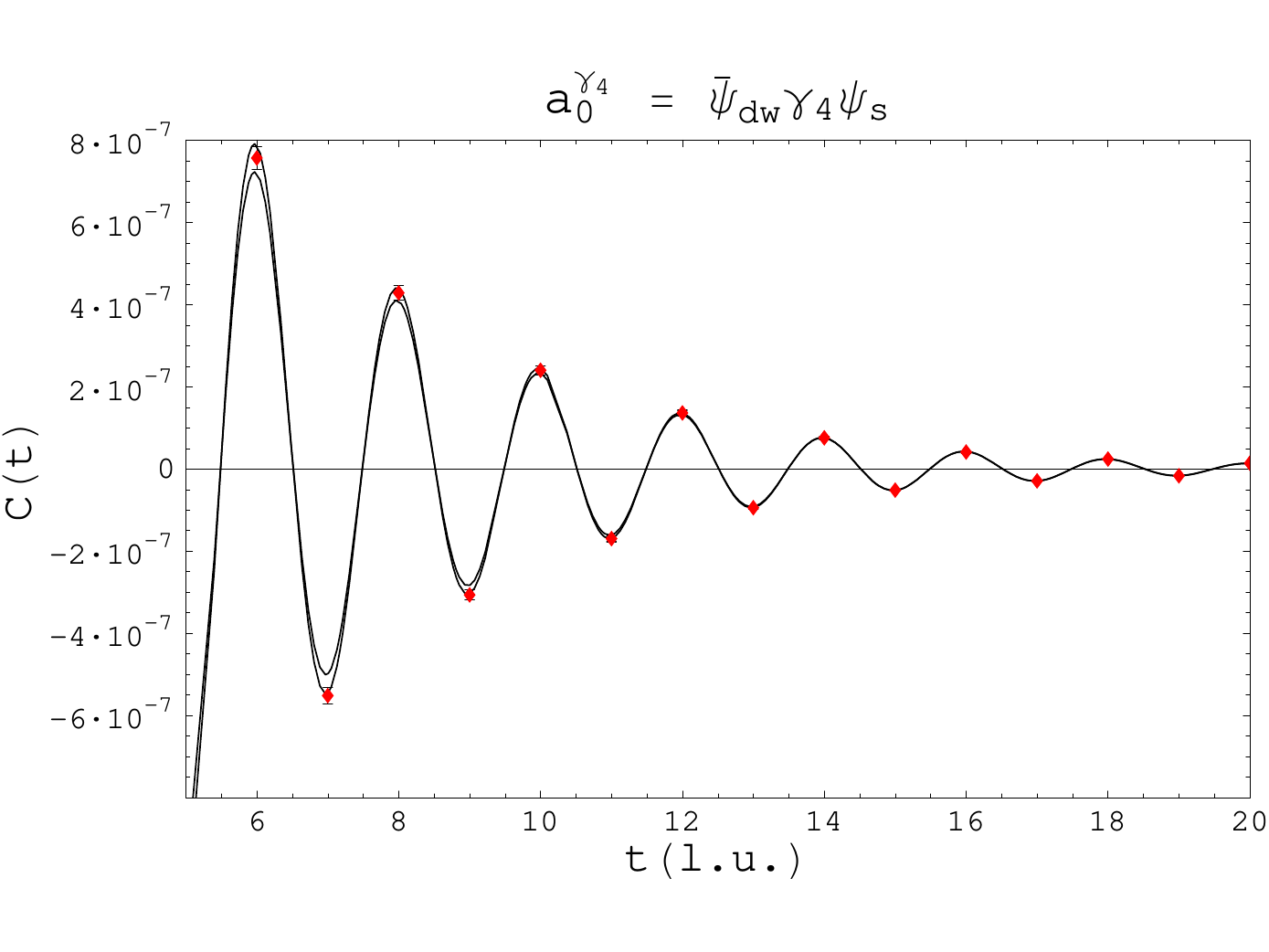}\\
$(a)$ & $(b)$
\end{tabular}
\caption{\label{fig:a02Corr} Correlation function created from a mixed $a_0^{\g_4}$ source.  In Figure~\ref{fig:a02Corr}$(a)$ we see the dramatic oscillatory effects which can occur when the oscillating amplitude is larger than the straight amplitude for degenerate or nearly degenerate masses.  In Figure~\ref{fig:a02Corr}$(b)$ we show two different fits to it described in App.~\ref{app:pisfromrhos}}
\end{figure}

In Table~\ref{tab:MIXEDresults}, we present the results of fits to the long time behavior of the various mixed meson sources with the full volume m010 propagators and compare them to the results of fitting the naive mixed pion source.  In Figure~\ref{fig:MvsPionCorr}, we display the best fit plots and results for the two-point correlation functions created with the ``pion" and ``rho" interpolating operators.   In Figure~\ref{fig:a02Corr} we plot the best fit and results for the correlation function created with the $\G = \g_4$ interpolating field, which demonstrates an even more dramatic example of the time-oscillatory nature of the mixed meson correlators.  In this case, the oscillating exponential has an amplitude roughly two orders of magnitude larger than the straight exponential as seen in Figure~\ref{fig:a02Corr}$(a)$.  In Figure~\ref{fig:a02Corr}$(b)$ we have actually plotted two different fits to the correlator, one with a forced degeneracy between the straight and oscillating state and the second letting the two mass parameters be independent, 
\begin{align}
	C_1(t) &= \Big[ A + B \textrm{cos}(\pi n_t) \Big] e^{-mt}\, ,
	\\
	C_2(t) &= A e^{-mt} + B \textrm{cos}(\pi n_t) e^{-Mt}\, ,
\end{align}
and we find consistent masses in both cases.
%
%
In Table~\ref{tab:MIXEDresults}, we collect the results of our determination of the mixed pion mass from all of the two-point correlation functions presented in this Appendix.
%
%
\begin{table}[t]
\caption{\label{tab:MIXEDresults} Mixed meson from $\rho$ and $a0$ and $b$ mesons.  We collect our results of fits to the long time behavior of the various two-point correlation functions created with the listed sources and compare the to that of the mixed pseudo scalar source.}
\begin{ruledtabular}
\begin{tabular}{cccccc}
Ensemble & source & $a\tilde{m}_\pi^{vs}$ & fit range & $\chi^2$ / d.o.f. & Q \\
\hline
m010(FV) & $\bar{\psi}_{dw}\, \g_5\, \psi_{s}$ & 0.2851(31) & 6--31& 20 / 23 & 0.65  \\
\hline
m010(FV) & $\bar{\psi}_{dw}\, \psi_{s}$ & 0.283(19) & 18--31 & 12 / 11 & 0.33 \\
m010(FV) & $\bar{\psi}_{dw}\, \g_4\, \psi_{s}$ &0.289(04) &8--31 & 17.6 / 21 & 0.68  \\
m010(FV) & $\bar{\psi}_{dw}\, \g_3\, \psi_{s}$ & 0.284(08) &9--31 & 17 / 20 & 0.68 \\
m010(FV) & $\bar{\psi}_{dw}\, \g_2\, \psi_{s}$ & 0.285(09) & 11--29 & 8.9 / 16 & 0.92 \\
m010(FV) & $\bar{\psi}_{dw}\, \g_1 \g_2\, \psi_{s}$ & 0.271(18) & 17--32 & 8.0 / 13 & 0.84 \\
m010(FV) & $\bar{\psi}_{dw}\, \g_3 \g_5\, \psi_{s}$ & 0.282(23) & 21--27 & 2.4 / 4 & 0.67
\end{tabular}
\end{ruledtabular}
\end{table}

%
%
\subsection{Mixed mesons and Dirichlet Boundary Conditions\label{app:MvsDBCs}}

In this appendix, we demonstrate the equivalence between the time shift symmetry and the taste-changing transformation which gives rise to the time-oscillatory states.  This makes it plausible that the Dirichlet boundary conditions, which break the time-shift symmetry, give rise to the observed splitting of the straight and oscillating states observed in the mixed meson correlation functions, Eq.~\eqref{eq:CvsDBC}.

In the continuum limit, one can associate the time oscillating field with a specific taste of staggered fermion field, most readily seen in momentum space.  Furthermore, using the momentum space time axis-reversal transformation for the staggered fermion fields~\cite{vandenDoel:1983mf,Golterman:1984cy}, one can show that under time-reversal, the staggered field undergoes an identical taste change as that from the time-doubling symmetry.  With time-anti periodic boundary conditions for the fermions, both of these transformations are symmetries of the staggered action, and therefore we conclude that the straight and oscillating states in the two-point correlation functions must be degenerate.

To show that the transformations of time-doubling and time axis-reversal correspond to the same taste change of the staggered field, it is simplest to work with the momentum space fermions.  To do this we construct a 16 component staggered field with components defined by the fermions at the different corners of the Brillouin zone,
\begin{equation}
	Q_g(k) = \chi(k +\pi_g)\, ,
\end{equation}
with $\pi_g$ defined in Eq.~\eqref{eq:pi_g}.  Then using the symmetries of the staggered action, one can show that under a time axis-flip, this fermion transforms as
\begin{equation}
	Q(k) \rightarrow \G_4 \G_5 \Xi_4 \Xi_5\, Q(k)\, ,
\end{equation}
where the set of $16\times 16$ matrices, $\G$ and $\Xi$ can be found for example in Refs.~\cite{vandenDoel:1983mf,Golterman:1984cy}.  If we begin with the staggered field in the central corner of the Brillouin zone, $Q_{0}(k)$, then one can show
\begin{equation}
	\G_4 \G_5 \Xi_4 \Xi_5\, Q_0(k) = Q_{4}(k)\, ,
\end{equation}
where $\pi_4 = \frac{\pi}{a} = (0,0,0,1)$.  From Eq.~\eqref{eq:stagTimeOsc}, we recognize that this is the exact taste change which occurs with the time-doubling symmetry.  Furthermore, if we apply this time axis-flip to all of the components of the staggered field given in Eq.~\eqref{eq:psi_s_h} which are related by the constraining equation, Eq.~\eqref{eq:stagFermSymm}, then one can show that the taste change from the time-doubling transformation and the time axis-flip transformation are identical.

This makes it plausible that the use of Dirichlet boundary conditions, which break this symmetry, give rise to the observed mass splitting between the straight and oscillating states observed in the mixed meson two point correlation functions.  We reiterate that this is not an issue for the mixed action calculations to date~\cite{Renner:2004ck,Bonnet:2004fr,Beane:2005rj,Edwards:2005kw,Edwards:2005ym,Beane:2006mx,Beane:2006pt,Beane:2006fk,Beane:2006kx,Alexandrou:2006mc,Beane:2006gj,Edwards:2006qx,Beane:2006gf}, as they only impose Dirichlet boundary conditions on the valence domain-wall fermions.  Our use of the Dirichlet boundary conditions for the staggered propagators used in the construction of the mixed meson two-point functions was one of convenience.

In Table~\ref{tab:MIXEDDBCs} we present our analysis of the quantity
\begin{equation}
	(a\tilde{M}_\pi^{vs})^2 - (a\tilde{m}_\pi^{vs})^2\, ,
\end{equation}
which shows no discernible dependence upon the pion mass.

%
%
\begin{table}[t]
\caption{\label{tab:MIXEDDBCs} Mixed meson and Dirichlet boundary conditions.  We present the mass splitting of the two mixed pions arising in a MA scheme with DBCs.  This mass splitting shows no dependence on the quark masses within our precision.}
\begin{ruledtabular}
\begin{tabular}{cclccc}
Ensemble & quantity & result & fit range & $\chi^2$ / d.o.f. & Q \\
\hline
m050(416 DBC) & $a\tilde{M}_\pi^{vs}$ & 0.532(3) & 6--13 & 4.95 / 4 & 0.292 \\
m050(416 DBC) & $a\tilde{m}_\pi^{vs}$ & 0.500(2) & 6--13 & 4.95 / 4 & 0.292 \\
m040(349 DBC) & $a\tilde{M}_\pi^{vs}$ &0.489(5) &6--13 &6.39 / 4 &0.172 \\
m040(349 DBC) & $a\tilde{m}_\pi^{vs}$ &0.457(2) &6--13 &6.39 / 4 &0.172 \\
m030(564 DBC) & $a\tilde{M}_\pi^{vs}$ &0.443(4) &6--12 &9.48 / 3 &0.02 \\
m030(564 DBC) & $a\tilde{m}_\pi^{vs}$ &0.409(2) &6--12 &9.48 / 3 &0.02 \\
m020(222 DBC)& $a\tilde{M}_\pi^{vs}$ & 0.393(6) & 6--12 & 1.64 / 3 & 0.650 \\
m020(222 DBC)& $a\tilde{m}_\pi^{vs}$ & 0.352(3) & 6--12 & 1.64 / 3 & 0.650 \\
m010(447 DBC) & $a\tilde{M}_\pi^{vs}$ &0.324(11) &6--12 &4.51 / 3 &0.21  \\
m010(447 DBC) & $a\tilde{m}_\pi^{vs}$ &0.282(3) &6--12 &4.51 / 3 &0.21 \\
\hline\hline
Ensemble & m010 & m020 & m030 & m040 & m050 \\
\hline
$(aM_\pi^{vs})^2 - (am_\pi^{vs})^2$ & 0.025(08) & 0.034(11) & 0.029(03) & 0.031(04) & 0.033(04) 
\end{tabular}
\end{ruledtabular}
\end{table}

\bibliography{%
bib_files/general,%
bib_files/EFT,%
bib_files/lattice_PQ,%
bib_files/lattice_FV,%
bib_files/NNEFT,%
bib_files/lattice_fermions,%
bib_files/lattice_physics,%
bib_files/lattice_general,%
bib_files/heavy_mesons,%
bib_files/lattice_QCD_spectrum,%
bib_files/hadron_structure}

\end{document}